\documentclass[]{IEEEtran}
\IEEEoverridecommandlockouts

\usepackage[pdftex]{graphicx}
\usepackage{epstopdf}
\usepackage[utf8]{inputenc} % allow utf-8 input
\usepackage[T1]{fontenc}    % use 8-bit T1 fonts
\usepackage{booktabs}       % professional-quality tables
\usepackage{amsfonts}       % blackboard math symbols
\usepackage{nicefrac}       % compact symbols for 1/2, etc.
\usepackage{microtype}      % microtypography
\newenvironment{proof}[1][Proof]{{\it #1. } }{\ \rule{0.5em}{0.5em}}
\usepackage{graphics, theorem, times, cite}
\usepackage{graphicx}
\usepackage{subfig}

\usepackage{tikz}
 
\usepackage[english]{babel}
\usepackage{ifpdf}
\usepackage{bm}

\ifpdf
	\usepackage[pdftex]{graphicx}
	\graphicspath{{./figures/}}
\else
	\usepackage[dvips]{graphicx}
	\graphicspath{./figures/}
\fi
\usepackage{color}
\usepackage{pgf, tikz, pgfplots, epstopdf}
\pgfplotsset{compat=1.18}
\usetikzlibrary{shapes, arrows, automata}
\usepackage{subfig}
\usepackage{amsmath}
\usepackage{amsfonts, amssymb,  bbm}
\usepackage{mathrsfs}
\usepackage{upgreek}
\usepackage{amsmath}
\DeclareMathOperator*{\argmax}{arg\,max}
\DeclareMathOperator*{\argmin}{arg\,min}
\usepackage{algorithm,algpseudocode}
\usepackage{enumerate}
\usepackage{multirow}
\usepackage{rotating}
\usepackage{graphicx}    % For \includegraphics
\usepackage{caption}     % For \caption in figures
\usepackage{subcaption}  % For \begin{subfigure}

\usepackage{needspace}

\input{mySymbol.sty}

\usepackage{booktabs}       % professional-quality tables
\usepackage{nicefrac}       % compact symbols for 1/2, etc.
\usepackage{microtype}      % microtypography
\usepackage{xcolor}         % colors

\newtheorem{proposition}{\hspace{0pt}\bf Proposition}
\newtheorem{pproposition}{\hspace{0pt}\bf Proposition}

\newtheorem{remark}{\hspace{0pt}\bf Remark}

\newcommand{\QED}{\hfill\ensuremath{\blacksquare}}

\definecolor{mutedblue}{RGB}{70, 130, 180}   % steel blue
\definecolor{mutedred}{RGB}{178, 34, 34}     % firebrick
\definecolor{mygreen}{RGB}{34, 139, 34}  

\usepackage[  
    colorlinks = true,       % Write links in color
    pdfborder  = {0 0 0},    % Do not use the distracting box around links
    urlcolor   = mutedblue,  % Standard URLs are written in dark blue 
    linkcolor  = mutedred,   % References to equations & sections are dark red
    citecolor  = mygreen]    % Citations are in green
   {hyperref}

\title{Long-Horizon Wireless Link Scheduling with State-Augmented Graph Neural Networks 
}

\author{Romina Garcia Camargo \quad Zhiyang Wang \quad Navid NaderiAlizadeh \quad Alejandro Ribeiro \thanks{RGC and AR are with the Department of Electrical and Systems Engineering, University of Pennsylvania, Philadelphia, PA (emails: \{rominag, aribeiro\}@seas.upenn.edu). ZW is with Halıcıoğlu Data Science Institute, UCSD, La Jolla, CA (email: zhw135@ucsd.edu). NN is with the Department of Biostatistics and Bioinformatics, Duke University, Durham, NC (email: navid.naderi@duke.edu). Preliminary results were presented in part at the Asilomar Conference on Signals, Systems, and Computers \cite{camargo2025wirelesslinkschedulingstateaugmented}}}

\usepackage[dvipsnames]{xcolor}

\newcommand{\rofs}[1]{#1 \odot \big[\, \mathbf{1} - \bbA#1    \,\big]_+}
\def \rs  {\rofs{\bbs}}
\def \rst {\rofs{\bbs(t)}}
\begin{document}
\pagestyle{plain}
\maketitle

\begin{abstract}
We address optimal link scheduling in large-scale wireless networks. The goal is to schedule transmissions over a time horizon so that to maximize sum rate while ensuring that average rates of each customer attain a minimum rate requirement. To this end, we formulate a constrained optimization problem and solve it using Lagrangian duality. Common primal-dual approaches lead to time invariant policies. Our constraint requires all links transmit a fraction of the time while avoiding interference, which calls for time-varying policies across time slots. We propose an iterative algorithm to sample optimal sequences of schedules and dual variables. The scheduling decisions are parameterized using a Graph Neural Network. We incorporate state-augmentation techniques to learn said parameterization, introducing dual variables as dynamic inputs to the policy. This augmentation enables the GNN to adapt scheduling decisions over time, balancing constraint satisfaction with performance maximization. We validate our approach through extensive numerical simulations, benchmarking against several baselines and considering varying constraint levels. %In particular, we investigate the model’s transferability, both theoretically and empirically, demonstrating its ability to generalize across different scales and transmission requirements. 
\end{abstract}


% \roms{define title}
\section{Introduction}
\label{sec:intro}

As wireless systems continue to grow in size and complexity, there is an increasing need for resource management algorithms that are both computationally efficient and capable of adapting to rapidly varying network conditions. One of the most fundamental challenges in large-scale wireless networks is link scheduling, which determines which links should transmit and when~\cite{shen2017fplinq, traditional3,naderializadeh2014itlinq,yi2015itlinq+,spatialdl, grlinq}. Effective link scheduling is essential for mitigating interference and ensuring fair and efficient utilization of scarce network resources.

Under certain interference models, link scheduling can be reduced to classic combinatorial problems such as edge coloring or the maximum independent set (MIS), both of which are known to be NP-hard \cite{npedgecoloring, npmis, shortestlinkscheduling, wlspowercontrol}. This inherent complexity has motivated extensive research, approached from both information-theoretic \cite{jafar2013topological,naderializadeh2014itlinq} and combinatorial \cite{wlsedgecoloring, efficienttdma, linkschedulingusinggnns} perspectives. Traditional methods have provided a strong foundation for analyzing network topology, modeling interference, and designing scheduling heuristics \cite{appralgforwls, appralg-exps, wlsphyinterf, geneticwls, shen2017fplinq}. 

More recently, learning-based approaches have gained traction for their ability to capture the intricacies of realistic and dynamic network environments \cite{lee2021graphembedding, g-mlriemannian, spatialdl, grlinq}. For example, \cite{spatialdl} introduces a deep neural network that bases scheduling decisions on geographic information, addressing interference while improving performance. Inspired by a related scenario, \cite{lee2021graphembedding} constructs graph embeddings from link distances, eliminating the need for precise channel state information. The work in \cite{g-mlriemannian} targets long-term network performance in dynamic (vehicular) settings, relying on a geometric machine learning approach grounded in Riemannian manifolds.%, which improves scalability and adaptability by leveraging the inherent geometric structure of wireless networks..
%Extending further, \cite{g-mlriemannian} presents a geometric machine learning approach grounded in Riemannian manifolds, which improves scalability and adaptability by leveraging the inherent geometric structure of wireless networks.%For example, \cite{spatialdl} introduces a deep neural network that bases scheduling decisions on geographic information, addressing interference while improving performance. Inspired by a related scenario, \cite{grlinq} proposes $K$-nearest-neighbor graphs to exploit local connectivity for scheduling, while \cite{lee2021graphembedding} constructs graph embeddings from link distances, eliminating the need for precise channel state information. Extending further, \cite{g-mlriemannian} presents a geometric machine learning approach grounded in Riemannian manifolds, which improves scalability and adaptability by leveraging the inherent geometric structure of wireless networks.

A common thread among many recent learning-based approaches is the use of graph-structured representations to encode network interactions. Wireless networks naturally admit a graph formulation, with nodes representing devices and edges capturing physical signal or interference relationships. This perspective has motivated the use of graph-based models, especially Graph Neural Networks (GNNs), for link scheduling~\cite{lee2021graphembedding, GBlinksbeams, linkschedulingusinggnns}. Under the primary interference model, two links interfere if they share a common user \cite{hajekb88}, leading to the construction of a conflict graph that explicitly captures these interactions \cite{firstconflictgraph}. This approach is used in \cite{linkschedulingusinggnns}, where link scheduling is reformulated as the problem of identifying an MIS in the conflict graph, with an emphasis on accelerating the computation of such sets. Nevertheless, most prior studies focus on maximizing the instantaneous sum rate \cite{lee2021graphembedding, linkschedulingusinggnns, spatialdl, GBlinksbeams, grlinq}, thereby overlooking the temporal coupling among scheduling decisions and failing to capture long-term performance metrics in dynamic wireless networks.

In this paper we propose a method for learning link scheduling policies that optimize rates averaged over a time horizon. Our specific formulation requires that all customers achieve a minimum rate requirement, with excess resources devoted to maximizing average sum rates (Section \ref{sec:problem}). It is noteworthy that the resulting constrained optimization problem is combinatorial in a very large dimensional space because the total number of variables involved is the product of the number of links and the number of slots in the time horizon. 

To \emph{learn} solutions of this constrained optimization problem we adopt state augmented algorithms \cite{calvofullana2023stateaugmentedconstrainedreinforcement, naderializadeh2022state}. These algorithms include an \emph{offline} training phase and and an \emph{online} execution phase. During the offline phase we train a learning parameterization to minimize a loss that penalizes for the violation of the constraints with penalty weights determined by a dual variable. During the online phase we consider time varying penalty weights and schedule links as dictated by the policy learned during the offline training phase. We complement the allocation of resources with an update of the penalty coefficients so that we increase weights for violated constraints and decrease weights for satisfied constraints (Section \ref{sec_state_augmented_learning}). We say that this is a state augmented algorithm because we can think of Lagrange multipliers as internal auxiliary states that guide link scheduling decisions. 

To explain why state augmented learning is a good idea we investigate link scheduling in the dual domain (Section \ref{sec:lagrangiandualdomain}). In particular, we consider dual gradient descent algorithms (Section \ref{sec_dual_gradient_descent}) and show that they yield schedule sequences that are approximately optimal in problems with large time horizons (Proposition \ref{prop:limits}). This is a remarkable fact because it implies that link scheduling problems over long time horizons are easier than expected. Approximate solutions can be found with an algorithm that is combinatorial with respect to the number of links but does \emph{not} have combinatorial complexity with respect to the length of the time horizon. State augmented learning works because it imitates dual gradient descent provided that the learning parameterization is trained to a small loss (Section \ref{sec_state_augmented_learning}). 

The choice of state augmented learning may seem contrived but it is justified because more natural choices have limitations (Remarks \ref{rmk_parameterization_unworkable} and \ref{rmk_parameterization_expensive}). These limitations are related to the fact that recovering optimal primal variables from optimal dual variables in the optimal link scheduling problem we consider here is impossible (Section \ref{sec_primal_recovery}).

We corroborate our theoretical findings with numerical experiments (Section \ref{sec:simu}). We show that the proposed algorithm returns schedules that satisfy the constraints for the vast majority of links (Section \ref{sec:hyp2-constraintsatisfaction}), such as the ones illustrated in Figure \ref{fig:illustration}. For under-scheduled links, the level of constraint violation is minor. We compare our method with several baselines and demonstrate that it achieves improved performance and faster runtime (Section \ref{sec:hyp3-baselines}). We further observe that the learned policy generalizes across different transmission requirements (Section \ref{sec:hyp4-differentdeltas}), adapts over time (Section~\ref{sec:hyp5-timeanalysis}), and achieves desirable effective rates (Section \ref{sec:hyp6-collisions}).

\section{Optimal Link Scheduling}
\label{sec:problem}
We address link scheduling in a device-to-device wireless network with $K$ links. Let $s_i\in\{0,1\}$ denote the status of link $i\in\{1,2,\dots, K\}$, where $s_i = 1$ indicates that the link $i$ is scheduled to transmit. Denote the scheduling status of all links by the vector $\bbs = [s_1; s_2; \cdots; s_K] \in \{0,1\}^K$. Following a primary interference model, we consider that two links interfere with each other if they share a common device \cite{hajekb88}. This interference model is described by a conflict graph $\bbG(\ccalV, \ccalE)$ in which the node set $\ccalV=\{1,2,\hdots, K\}$ represents different links and the edge set $\ccalE$ includes the edge $(i, j)$ whenever links $i$ and $j$ interfere \cite{firstconflictgraph}. The adjacency matrix $\bbA\in\{0,1\}^{K\times K}$ of this conflict graph is a sparse matrix with nonzero entries $A_{ij} = 1$ if and only if $(i,j)$ is in the edge set $\ccalE$. 

With these definitions we can indicate \emph{successful} transmissions $\bbr(\bbs)$ of a schedule $\bbs$ by computing the Hadamard product between schedules $\bbs$ and the nonnegative projection $[\mathbf{1}  - \bbA \bbs]_+$,
\begin{align}\label{eqn_rate_pointwise}
    \bbr(\bbs) = \rs .
\end{align}
Indeed, in the conflict graph adjacency we have $A_{ij} = 1$ whenever link $j$ \emph{can} interfere with link $i$. Since we also have that $s_j=1$ whenever $j$ is scheduled, the product $A_{ij} s_j = 1$ indicates that link $s_j$ \emph{is} actually interfering with link $i$. For a given schedule $\bbs$ the entries $(\bbA \bbs)_i = \sum_{j} A_{ij} s_j$ %of the product $\bbA \bbs$ 
count the number of scheduled links $j$ that interfere with link $i$. We thus have $1 - (\bbA \bbs)_i = 1$ only when no other link $j$ is interfering with link $i$. If this link is scheduled, the transmission is successful. %In \eqref{eqn_rate_pointwise} this implies that  $\bbr_i(\bbs) = 1$ because $s_i=1$ and $1 - (\bbA \bbs)_i = 1$ as we have already explained. 
The nonnegative projection $[\mathbf{1}  - \bbA \bbs]_+$ is needed to make $\bbr(\bbs)$ a proper indicator variable with entries $\bbr_i(\bbs) \in \{0,1\}$.

\begin{figure*}

\centering

\includegraphics[width=0.25\linewidth]{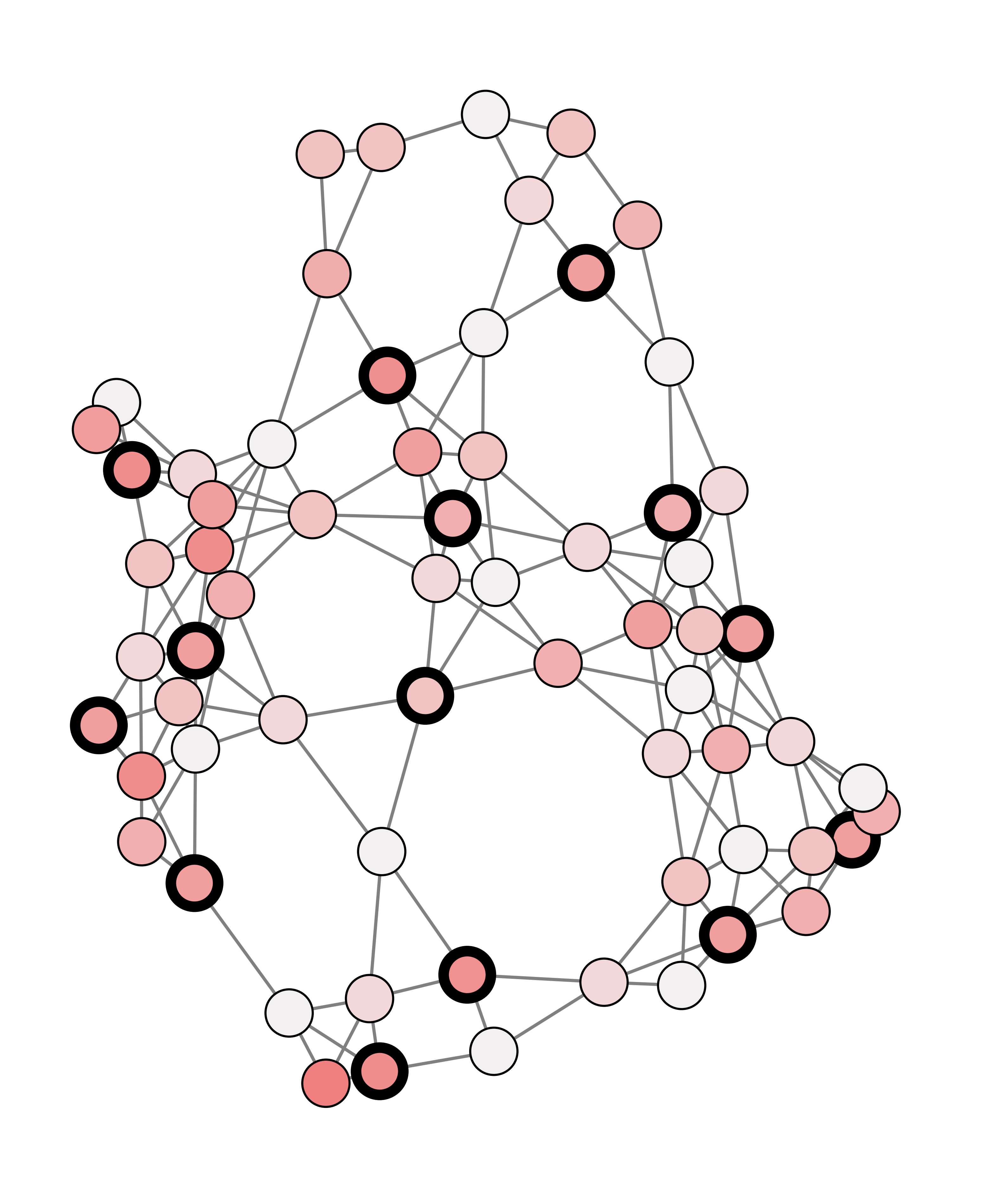}
\includegraphics[width=0.25\linewidth]{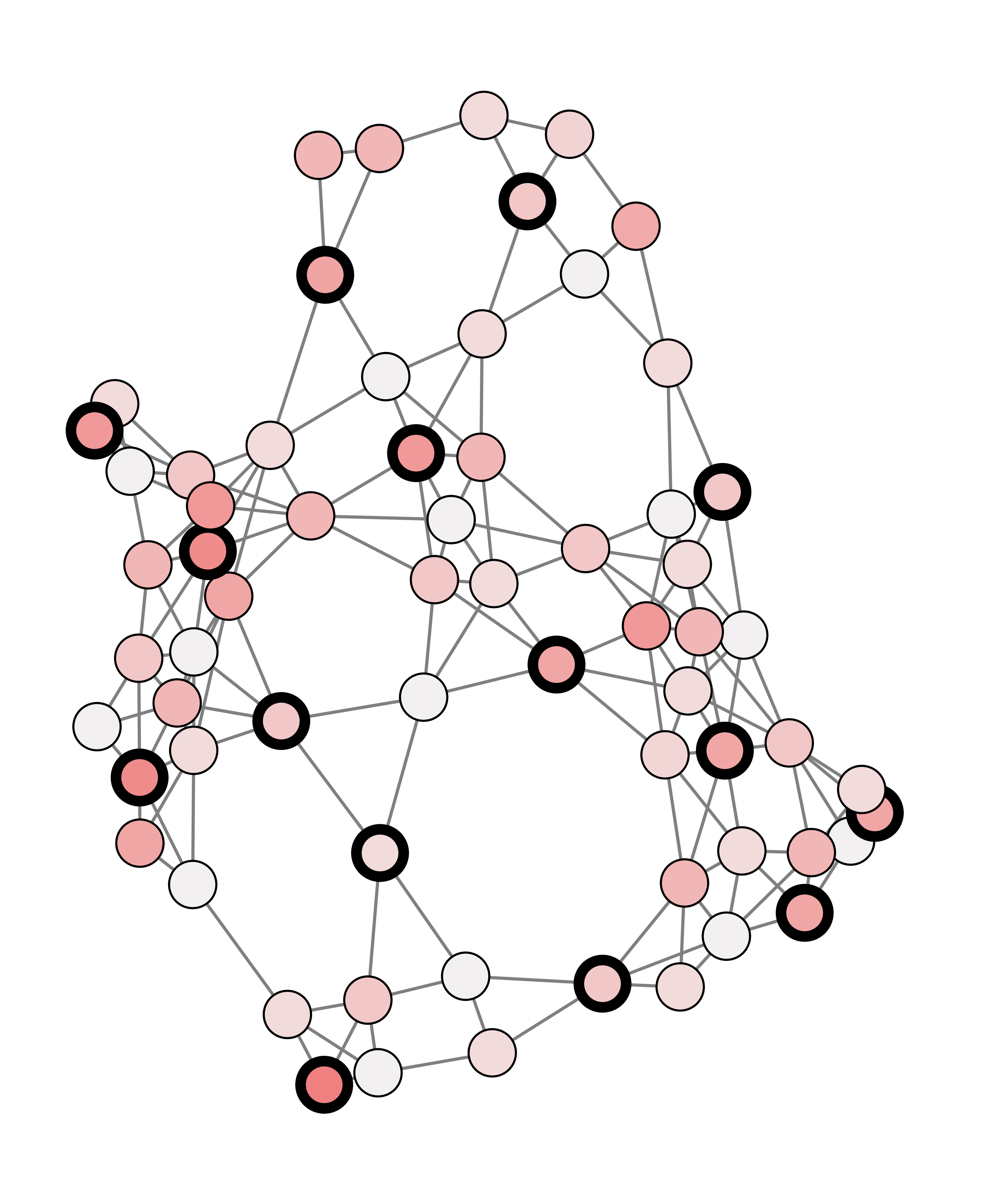}
\includegraphics[width=0.3\linewidth]{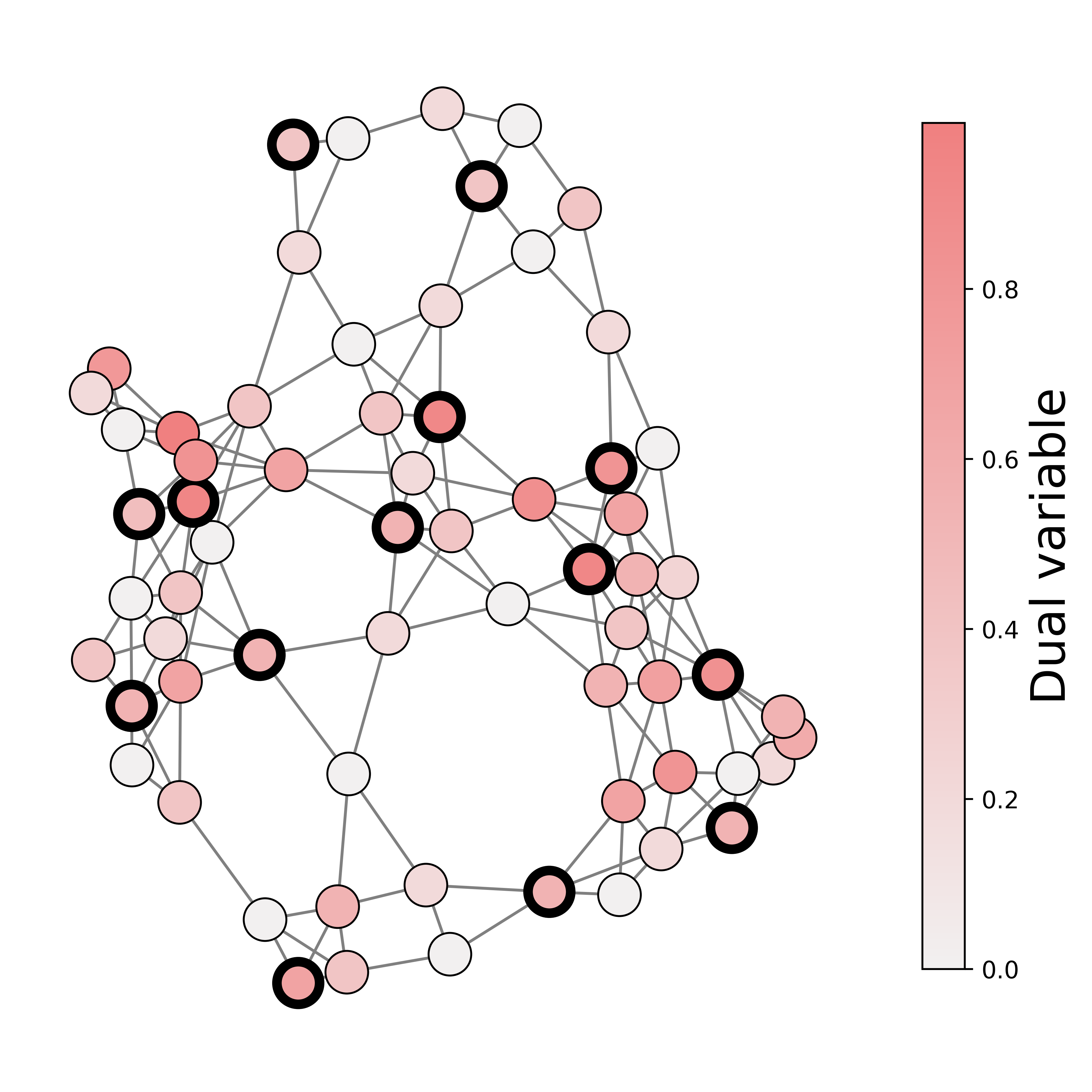}

\caption{Illustration of an optimal policy for different time steps. Links scheduled for transmission are highlighted with a thicker black line. The changes in the dual variable penalizing constraint violations are shown in shades of red. Only a portion of the graph is displayed for enhanced clarity. The scheduling decisions effectively alternate between non-interfering links, and links with higher constraint violation levels are scheduled.}

\label{fig:illustration}

\end{figure*}

Since end users experience Quality of Service (QoS) over several transmission slots, the average rate over a time horizon $T$ is of interest.  We therefore introduce $\bbs(t)$ to indicate the schedule for time $t$ and $\bar\bbr$ to denote the rate of successful transmissions averaged over the time horizon,
\begin{align}\label{eqn_rate_average}
    \bar\bbr [\bbs(1: T)] 
        = \frac{1}{T} \sum_{t=1}^{T} \bbr[\bbs(t)]
        = \frac{1}{T} \sum_{t=1}^{T} \rst,
\end{align}
The objective of wireless link scheduling is to design a policy that satisfies the interference constraints dictated by the network topology, while maximizing the long-term sum rate $\mathbf{1}^\top\bar\bbr$. We restrict our attention to scheduling sequences %$\{\bbs(t)\}_{t=0}^{T-1}$ 
$\bbs(1: T)$ that meet a minimum transmission requirement for each link, specified by the vector $\bm\Delta \in \reals_+^{K}$, such that $\bar\bbr \geq \mathbf{\Delta}$. With these definitions, the optimal scheduling problem can be formulated as the constrained optimization problem,
 \begin{alignat}{3} \label{eq:wls}
     \bbs^\star(1: T) 
         ~=~ & \argmax_{ \bbs(t)\in\{0,1\}^K} ~  
                   && \frac{1}{T} \sum_{t=1}^{T}  \bbone^\top \rst  , 
                            \nonumber \\
             & \text{s.t.}  
                   && \frac{1}{T} \sum_{t=1}^{T} \rst
                            \geq \bbDelta,
 \end{alignat}
where we call $P_T$ the optimal long-term sum rate achieved with $\bbs^\star(1: T)$. Note that the minimum transmission constraint implies that each link $i$ transmits a fraction of the time $\bm{\Delta}_i$. To meet this requirement while avoiding interference we require that policies are time-dependent, as illustrated in Figure \ref{fig:illustration}. 

%%%%%%%%%%%%%%%%%%%%%%%%%%%%%%%%%%%%%%%%%%%%%%%%%%%%%%%%%%%%%%%%%%%%%%%%%%%%%%%%
%%%   F   I   G   U   R   E   %%%%%%%%%%%%%%%%%%%%%%%%%%%%%%%%%%%%%%%%%%%%%%%%%%
%%%%%%%%%%%%%%%%%%%%%%%%%%%%%%%%%%%%%%%%%%%%%%%%%%%%%%%%%%%%%%%%%%%%%%%%%%%%%%%%

\section{Link Scheduling in the Dual Domain}
\label{sec:lagrangiandualdomain}
To derive optimal scheduling policies, we introduce a penalty version with the Lagrangian dual. The Lagrangian of \eqref{eq:wls} is a linear combination of the objective and constraints in which the components $\bm{\lam}_i$ of a dual variable $\bm{\lambda}$ define the weights of the rate constraints $\bar{\bbr}_i [\bbs(1:T)] \geq \bm{\Delta}_i$. Formally, introduce a dual variable $\bm{\lambda}$ and consider an \emph{instantaneous} resource allocation $\bbs$ to define the \emph{instantaneous} Lagrangian
\begin{align}\label{eqn_time_t_lagrangian}
    \ccalM(\bbs, \bblam) 
        & = \bbone^\top  \Big[ \bbs \odot \big[ \mathbf{1} - \bbA\bbs \big]_+ \Big] 
               + \bblam^\top  \Big[ \bbs \odot \big[ \mathbf{1} - \bbA\bbs \big]_+ \! - \bbDelta \Big] 
                     \nonumber \\ 
        & = (\bbone + \bblam)^\top  \Big[\, \rs  \, \Big] - \bblam^\top\bbDelta .
\end{align}
The maximization of $\ccalM(\bbs, \bblam)$ differs from \eqref{eq:wls} in two ways: (i) We consider a single resource allocation $\bbs$ instead of a sequence of resource allocations $\bbs(1:T)$. (ii) Rate constraints are not enforced but rather penalized with weight coefficients $\bblam_i$ determined by the entries of the dual variable $\bblam$.

We can think of the instantaneous Lagrangian in \eqref{eqn_time_t_lagrangian} as the Lagrangian of \eqref{eq:wls} for the particular case of $T=1$. For generic $T$, all of the summands that appear in the objective and constraints of \eqref{eq:wls} are evaluated at different schedule variables $\bbs(t)$ but have the same functional form. Thus, the Lagrangian of \eqref{eq:wls} can be written as a time average of instantaneous Lagrangians $\ccalM(\bbs (t), \bblam)$, evaluated with a common multiplier $\bblam$ and different schedules $\bbs(t)$,
\begin{align}\label{eq:lag}
    \ccalL \Big(\, \bbs(1: T), \bblam \, \Big) 
        = \frac{1}{T} \sum_{t=1}^{T} 
              \ccalM(\bbs (t), \bblam).
\end{align}
The Lagrangian in \eqref{eq:lag} is a penalty formulation of \eqref{eq:wls}. Instead of requiring schedules to achieve rates $\bm{\bar{\bbr}}_i [\bbs(1:T)] \geq \bm{\Delta}_i$, we penalize the violation of the rate constraints according to the values of the entries $\bm{\lam}_i$ of the dual variable $\bblam$. Thus, in lieu of the optimal primal variables $\bbs^\star(1: T)$ we settle for the Lagrangian maximizing variables $\bbs^\dagger(1: T, \bblam)$ computed as
\begin{align}\label{eq:maxofL}
    \bbs^\dagger(1:T, \bblam) 
        = \argmax_{\bbs(1:T)\in\{0,1\}^K} 
              \ccalL \Big(\, \bbs(1:T),\, \bblam \, \Big).
\end{align}
It is important for forthcoming discussions that the Lagrangian maximizing variable $\bbs^\dagger(1:T, \bblam)$ is a set. There may be -- indeed, most likely there are -- values of $\bblam$ for which this set contains more than one value. We will use the symbol $\bbs^\dagger(1: T, \bblam)$ to also denote elements of the set of Lagrangian maximizers with context clarification.

Lagrangian values associated with Lagrangian maximizers define the dual function $d(\bblam) = \ccalL \left(\bbs^\dagger(1: T, \bblam), \bblam\right)$ -- if there is more than one Lagrangian maximizer they all achieve the same value. It is ready and well known that $d(\bblam)$ is an upper bound on the optimal yield. I.e. that $P_T \leq d(\bblam)$ for any multiplier $\bblam$ \cite{boyd2004convex}. It is therefore sensible to focus on the multiplier $\bblam^\star$ that minimizes this bound. This is the optimal dual variable which we define as,
\begin{align}
  \bblam^\star 
      ~=~ \argmin_{\bblam\in \reals_+^K} d(\bblam)
      ~=~ \argmin_{\bblam\in \reals_+^K} \ccalL \Big(\, \bbs^\dagger(1:T, \bblam),\, \bblam \, \Big).
\end{align}
Since $\bblam^\star$ yields the dual function value $D_T = d(\bblam^\star)$ closest to the optimal primal yield $P_T$, we can adopt the Lagrangian maximizers $\bbs^\dagger(1:T, \bblam^\star)$ as an alternative to the optimal schedule $\bbs^\star(1:T)$. I.e., we redefine our goal as the determination of the Lagrangian maximizing schedules,
\begin{align}\label{eqn_lagrangian_maximizers_optimal}
    \bbs^\dagger(1:T, \bblam^\star) 
        = \argmax_{\bbs(1:T)\in\{0,1\}^K} 
              \ccalL \Big(\, \bbs(1:T),\, \bblam^\star \, \Big).
\end{align}
This is appealing for learning because training a parametrization to estimate solutions of \eqref{eqn_lagrangian_maximizers_optimal} over a distribution of network realizations is a standard learning problem. The added complexity of determining the optimal multiplier $\bblam^\star$ is not prohibitive because the dual function is always convex \cite{boyd2004convex}.

The challenge with using the Lagrangian maximizers $\bbs^\dagger(1:T, \bblam^\star)$ in \eqref{eqn_lagrangian_maximizers_optimal} in lieu of the optimal schedule $\bbs^\star(1:T)$ [cf. \eqref{eq:wls}] is that the properties of $\bbs^\dagger(1:T, \bblam^\star)$  make it unworkable as a proxy for $\bbs^\star(1:T)$. The solution to this challenge is to leverage dual gradient descent dynamics. We explain why $\bbs^\dagger(1:T, \bblam^\star)$ is an unworkable proxy for $\bbs^\star(1:T)$ in Section \ref{sec_primal_recovery} and how dual gradient descent dynamics circumvent this challenge in Section \ref{sec_dual_gradient_descent}. We leverage the latter to introduce the state augmented learning as an imitation of dual gradient descent(Section \ref{sec_state_augmented_learning}).

%%%%%%%%%%%%%%%%%%%%%%%%%%%%%%%%%%%%%%%%%%%%%%%%%%%%%%%%%%%%%%%%%%%%%%%%%%%%%%%%
%%%   S   E   C   T   I   O   N   %%%%%%%%%%%%%%%%%%%%%%%%%%%%%%%%%%%%%%%%%%%%%%
%%%%%%%%%%%%%%%%%%%%%%%%%%%%%%%%%%%%%%%%%%%%%%%%%%%%%%%%%%%%%%%%%%%%%%%%%%%%%%%%

\subsection{Primal Infeasibility of Lagrangian Maximizers}\label{sec_primal_recovery}

According to \eqref{eq:lag}, the Lagrangian of \eqref{eq:wls} is a sum of instantaneous Lagrangians $\ccalM(\bbs (t), \bblam)$. It is notable that only the variable $\bbs(t)$ appears in each instantaneous Lagrangian. Thus, to maximize the (aggregate) Lagrangian $\ccalL(\bbs(1:T), \bblam)$ we can maximize each instantaneous Lagrangian $\ccalM(\bbs (t), \bblam)$ separately. Also notable, instantaneous Lagrangians for different times are identical. They differ only in that the optimization variables are different, but they all share the same parameters -- the conflict graph $\bbA$, the multiplier $\bblam$ and the rate requirement $\bbDelta$. Thus, the schedules $\bbs^\dagger(t, \bblam)$ that maximize $\ccalM(\bbs(t), \bblam)$, and $\ccalL \left(\bbs(1: T), \bblam\right)$ by extension, are the same for all $t$. We summarize these two facts in the following proposition.

%%%%%%%%%%%%%%%%%%%%%%%%%%%%%%%%%%%%%%%%%%%%%%%%%%%%%%%%%%%%%%%%%%%%%%%%%%%%%%%%
%%%   P   R   O   P   O   S   I   T   I   O   N   %%%%%%%%%%%%%%%%%%%%%%%%%%%%%%
%%%%%%%%%%%%%%%%%%%%%%%%%%%%%%%%%%%%%%%%%%%%%%%%%%%%%%%%%%%%%%%%%%%%%%%%%%%%%%%%

\begin{proposition}\label{prop_lagrangian_maximizers}
Let $\bbs^\dagger(1:T, \bblam)$ be a maximizing \eqref{eq:maxofL} schedule of the Lagrangian \eqref{eq:lag} and let $\bbs^\dagger(t, \bblam)$ be the time $t$ entry of this schedule. Time $t$ entries can be computed by maximizing the instantaneous Lagrangian \eqref{eqn_time_t_lagrangian},  
\begin{align}\label{eq:maxofLtime}
    \bbs^\dagger(t, \bblam) 
        ~\equiv~ \argmax_{ \bbs \in \{0,1\}^K } 
                     \ccalM (\bbs, \bblam ).
\end{align}
Consequently, the set of time $t$ and time $v$ Lagrangian maximizers contain the same schedules. I.e.,
\begin{align}\label{eq:maxofLtime-tu}
    \bbs^\dagger(t, \bblam) ~\equiv~ \bbs^\dagger(v, \bblam),
\end{align}
for all dual variables $\bblam$ and all time pairs $t$ and $v$.

\end{proposition}

%%%%%%%%%%%%%%%%%%%%%%%%%%%%%%%%%%%%%%%%%%%%%%%%%%%%%%%%%%%%%%%%%%%%%%%%%%%%%%%%
%%%   P   R   O   O   F   %%%%%%%%%%%%%%%%%%%%%%%%%%%%%%%%%%%%%%%%%%%%%%%%%%%%%%
%%%%%%%%%%%%%%%%%%%%%%%%%%%%%%%%%%%%%%%%%%%%%%%%%%%%%%%%%%%%%%%%%%%%%%%%%%%%%%%%

\begin{proof}
    See Appendix \ref{sec:proposition}. 
\end{proof}

%%%%%%%%%%%%%%%%%%%%%%%%%%%%%%%%%%%%%%%%%%%%%%%%%%%%%%%%%%%%%%%%%%%%%%%%%%%%%%%%
%%%   M   A   I   N       M   A   T   T   E   R   %%%%%%%%%%%%%%%%%%%%%%%%%%%%%%
%%%%%%%%%%%%%%%%%%%%%%%%%%%%%%%%%%%%%%%%%%%%%%%%%%%%%%%%%%%%%%%%%%%%%%%%%%%%%%%%

Proposition \ref{prop_lagrangian_maximizers} holds for any multiplier $\bblam$, optimal multipliers $\bblam^\star$ in particular. In this latter case, Result \eqref{eq:maxofLtime} is positive. It indicates that the maximization of the Lagrangian \eqref{eqn_lagrangian_maximizers_optimal} decomposes into $T$ separate maximizations of instantaneous Lagrangians. This implies that to find $\bbs^\dagger(1:T, \bblam^\star)$ we need to solve $T$ subproblems that are combinatorial in $K$ variables. In contrast to the primal problem \eqref{eq:maxofL} which is combinatorial in $KT$ variables, this is a significant improvement of computational tractability.

Result \eqref{eq:maxofLtime} is also negative. It indicates that the Lagrangian maximizing schedule $\bbs^\dagger(1:T, \bblam^\star)$ is such that $\bbs^\dagger(t, \bblam^\star) = \bbs^\dagger(v, \bblam^\star)$ for all pairs of times $t$ and $v$. This property makes it an unsuitable candidate for an approximate solution of the primal problem \eqref{eq:maxofL}. Indeed, a schedule $\bbs(t)$ with nonzero rate $\textbf{r}_i(\bbs(t)) > 0$ in link $i$ \emph{requires} zero rates $\textbf{r}_j(\bbs(t)) = 0$ in adjacent links in the conflict graph [cf. \eqref{eqn_rate_pointwise}]. Thus, except for situations in which rate requirements are $\bm{\Delta}_j = 0$ for some nodes, schedules in which $\bbs(t) = \bbs(v)$ for all $t$ and $v$ are not feasible in \eqref{eq:wls}. To satisfy requirements we need schedules that change over time so that different links can transmit at different times without conflicting with each other. Thus, if Lagrangian maximizers $\bbs^\dagger(1:T, \bblam^\star)$ are such that $\bbs^\dagger(t, \bblam^\star) = \bbs^\dagger(v, \bblam^\star)$ we have that the Lagrangian maximizing schedules in \eqref{eqn_lagrangian_maximizers_optimal} are not even feasible in \eqref{eq:wls} -- not to mention (close to) optimal.

The unworkability of the Lagrangian maximizing schedule $\bbs^\dagger(1:T, \bblam^\star)$ as a solution of \eqref{eq:wls} requires a more creative use of duality for learning solutions of \eqref{eq:wls}. This can be done by leveraging properties of dual gradient descent dynamics that we discuss next (Section \ref{sec_dual_gradient_descent}).    

%%%%%%%%%%%%%%%%%%%%%%%%%%%%%%%%%%%%%%%%%%%%%%%%%%%%%%%%%%%%%%%%%%%%%%%%%%%%%%%%
%%%   R   E   M   A   R   K   %%%%%%%%%%%%%%%%%%%%%%%%%%%%%%%%%%%%%%%%%%%%%%%%%%
%%%%%%%%%%%%%%%%%%%%%%%%%%%%%%%%%%%%%%%%%%%%%%%%%%%%%%%%%%%%%%%%%%%%%%%%%%%%%%%%

\begin{remark} \normalfont

A caveat to the infeasibility of the Lagrangian maximizing schedules in \eqref{eqn_lagrangian_maximizers_optimal} is that \eqref{eq:maxofLtime} claims the equivalence of Lagrangian maximizing \emph{sets.} It is then possible that when  $\bblam=\bblam^\star$, the Lagrangian maximizing sets $\bbs^\dagger(t, \bblam) = \bbs^\dagger(t, \bblam^\star)$ contain multiple schedules. In this case we can select different schedules $\bbs^\dagger(t, \bblam^\star) \neq \bbs^\dagger(v, \bblam^\star)$ at different times to obtain a feasible joint schedule $\bbs^\dagger(1:T, \bblam^\star)$. Having Lagrangian maximizing sets $\bbs^\dagger(t, \bblam^\star)$ with multiple elements is all but guaranteed but choosing the right combination of schedules from this set is, in general, as difficult as solving \eqref{eq:wls}. %\red{See Appendix \ref{??} for details.} \blue{Alejandro is not sure that this appendix is worth writing.}

\end{remark}

%%%%%%%%%%%%%%%%%%%%%%%%%%%%%%%%%%%%%%%%%%%%%%%%%%%%%%%%%%%%%%%%%%%%%%%%%%%%%%%%
%%%   S   E   C   T   I   O   N   %%%%%%%%%%%%%%%%%%%%%%%%%%%%%%%%%%%%%%%%%%%%%%
%%%%%%%%%%%%%%%%%%%%%%%%%%%%%%%%%%%%%%%%%%%%%%%%%%%%%%%%%%%%%%%%%%%%%%%%%%%%%%%%

\subsection{Dual Gradient Descent} \label{sec_dual_gradient_descent}

Dual gradient descent algorithms follow negative subgradients of the dual function towards optimal dual variables $\bblam^\star$. Formally, consider an iteration index $u$ and let $\bblam(u)$ be the corresponding dual iterate. At iteration $u$ we determine a schedule $\bbs^\ddagger(u)$ that maximizes the \emph{instantaneous} Lagrangian $\ccalM(\bbs, \bblam(u))$ associated with this multiplier,
\begin{alignat}{3}\label{eqn_dga_maximizers}
    \bbs^\ddagger(u) 
        & ~\in~ && \argmax_{ \bbs \in \{0,1\} }  
                    ~ \ccalM\Big(\, \bbs, \bblam(u) \, \Big)  \nonumber\\ 
        & ~  =~ && \argmax_{ \bbs \in \{0,1\} }
                    ~ (\bbone + \bblam(u))^\top  \Big[\, \rs  \, \Big] .
\end{alignat}
In the second equality we do not write the term $\bblam^\top\bbDelta$ because it does not depend on the schedule choice -- and therefore does not factor in finding the maximizing schedule $\bbs^\ddagger(u)$.

Because of Proposition \ref{prop_lagrangian_maximizers} we know that a schedule with $\bbs(t) = \bbs^\ddagger(u)$ for all $t$ -- recall that $u$ and $t$ represent different types of indices; while index $u$ is a dual iteration step, index $t$ is a scheduling slot -- maximizes the (average) Lagrangian $\ccalL(\bbs(1 : T), \bblam)$. Because we also know that evaluating constraint slacks at Lagrangian maximizers yields a subgradient of the dual function \cite{bertsekas1999nonlinear}, we use the maximizers in \eqref{eqn_dga_maximizers} to construct a subgradient of the dual function. A concrete expression is given in the following proposition. 

%%%%%%%%%%%%%%%%%%%%%%%%%%%%%%%%%%%%%%%%%%%%%%%%%%%%%%%%%%%%%%%%%%%%%%%%%%%%%%%%
%%%   P   R   O   P   O   S   I   T   I   O   N   %%%%%%%%%%%%%%%%%%%%%%%%%%%%%%
%%%%%%%%%%%%%%%%%%%%%%%%%%%%%%%%%%%%%%%%%%%%%%%%%%%%%%%%%%%%%%%%%%%%%%%%%%%%%%%%

\begin{proposition}\label{prop_subgradient}

The instantaneous constraint slack associated with the instantaneous Lagrangian maximizer $\bbs^\ddagger(u)$ \eqref{eqn_dga_maximizers} is a subgradient of the dual function $d(\bblam)$ at $\bblam = \bblam(u)$. I.e., the vector 
\begin{align}\label{eqn_subgradient}
    \bbg(\bblam(u)) ~=~   \rofs{\bbs^\ddagger(u)} ~-~ \bbDelta
\end{align}
is such that $\bbg^\top(\bblam(u))(\bblam(u) - \bblam^\star) \geq d(\bblam(u)) - d(\bblam^\star) \geq 0$.

\end{proposition}

%%%%%%%%%%%%%%%%%%%%%%%%%%%%%%%%%%%%%%%%%%%%%%%%%%%%%%%%%%%%%%%%%%%%%%%%%%%%%%%%
%%%   P   R   O   O   F   %%%%%%%%%%%%%%%%%%%%%%%%%%%%%%%%%%%%%%%%%%%%%%%%%%%%%%
%%%%%%%%%%%%%%%%%%%%%%%%%%%%%%%%%%%%%%%%%%%%%%%%%%%%%%%%%%%%%%%%%%%%%%%%%%%%%%%%

\begin{proof} See Appendix \ref{sec:proposition}. \end{proof}

%%%%%%%%%%%%%%%%%%%%%%%%%%%%%%%%%%%%%%%%%%%%%%%%%%%%%%%%%%%%%%%%%%%%%%%%%%%%%%%%
%%%   M   A   I   N       M   A   T   T   E   R   %%%%%%%%%%%%%%%%%%%%%%%%%%%%%%
%%%%%%%%%%%%%%%%%%%%%%%%%%%%%%%%%%%%%%%%%%%%%%%%%%%%%%%%%%%%%%%%%%%%%%%%%%%%%%%%

The subgradient expression in Proposition \ref{prop_subgradient} is unusual in that we use a schedule that maximizes the \emph{instantaneous} Lagrangian [cf. \eqref{eqn_dga_maximizers}] to find a descent direction of the dual function $d(\bblam)$ obtained from maximization of the \emph{average} Lagrangian [cf. \eqref{eq:maxofL}]. Unusual as it may, it still provides a descent direction for the dual function $d(\bblam)$. We therefore introduce a stepsize $\eta$ %\blue{($\eta$ for dual and $\zeta$ for primal, $\zeta$ comes right before $\eta$ in the greek alphabet)}
and proceed to update the dual variable as
\begin{align}\label{eq:lambdaupdate}
    \bblam(u+1) 
        &~=~ \Big[\, \bblam(u) - \eta  \, \bbg(\bblam(u)) \,\Big]_+  \\ \nonumber 
        &~=~ \Big[\bblam(u) - \eta  \Big[\rofs{\bbs^\ddagger(u)} - \bbDelta \Big] \Big]_+ 
\end{align}
Since the dual function is convex, this iterative update is guaranteed to approach the set of optimal multipliers $\bblam^\star$. Asymptotically, we are guaranteed to be in the neighborhood of size $\ccalO(\eta)$ of some optimal multiplier $\bblam^\star$ (\cite{bertsekas1999nonlinear}, see Appendix \ref{prop_bound}). 

Since dual iterates $\bblam(u)$ approach $\bblam^\star$, Lagrangian maximizers $\bbs^\ddagger(u)$ approach maximizers associated with optimal dual variables $\bbs^\dagger(\bblam^\star)$. However, as we already argued in Section \ref{sec_primal_recovery}, the latter are \emph{not} workable as proxies for the optimal schedules $\bbs^\star(1:T)$. Nevertheless, the \emph{sequence} of Lagrangian maximizers does have an interesting property that makes it workable as a solution to the optimal scheduling problem \eqref{eq:wls}. To state this result define the average rate,
\begin{align}\label{eqn_trajectory_rates}
    \bar\bbr\big[\bbs^\ddagger(1: T)\big] 
        ~=~ \frac{1}{T} \sum_{u=1}^{T} \rofs{\bbs^\ddagger(u)} .
\end{align}
This is the average rate that we achieve if we execute the \emph{sequence} $\bbs^\ddagger(1: T)$ made up of Lagrangian maximizing schedules $\bbs^\ddagger(u)$ associated with multipliers $\bm{\lam}(u)$ that make up a \emph{sequence} of multipliers $\bblam(1: T)$ generated by running the dual subgradient updates in \eqref{eq:lambdaupdate}. 

% It can be shown that the average rates in \eqref{eqn_trajectory_rates} are feasible and nearly optimal in \eqref{eq:wls}. This is a result from \cite{navidsa,ribeiro2010ergodic} that we repeat next for completeness.

%%%%%%%%%%%%%%%%%%%%%%%%%%%%%%%%%%%%%%%%%%%%%%%%%%%%%%%%%%%%%%%%%%%%%%%%%%%%%%%%
%%%   P   R   O   P   O   S   I   T   I   O   N   %%%%%%%%%%%%%%%%%%%%%%%%%%%%%%
%%%%%%%%%%%%%%%%%%%%%%%%%%%%%%%%%%%%%%%%%%%%%%%%%%%%%%%%%%%%%%%%%%%%%%%%%%%%%%%%

\begin{proposition}\label{prop:limits}
    Let $\bar\bbr[\bbs(1: T)]$ be the average rates in \eqref{eqn_trajectory_rates} obtained by executing the sequence of Lagrangian maximizing schedules $\bbs^\ddagger(\bblam(u))$. These maximizers are associated with the sequence of multipliers $\bblam(1: T)$ obtained from running the subgradient updates in \eqref{eq:lambdaupdate}. The average rates $\bar\bbr[\bbs(1: T)]$ are asymptotically nearly optimal
\begin{align}\label{eq:optimality}
\bbone^\top\bar\bbr\big[\bbs^\ddagger(1: T)\big]
            \geq P^\star_T - \frac{\eta}{2}\|\bbM-\bbDelta\|^2-\frac{|\bblam(1)|^2}{2\eta T},
\end{align}
and asymptotically feasible, 
\begin{align}\label{eq:feasibility}
        \bar\bbr\big[\bbs^\ddagger(1: T)\big]
            \geq \mathbf{\Delta}- %\frac{\bblam(T)-\bblam(1)}{\eta T}.
            \frac{\bblam(T)}{\eta T}.
\end{align}
In \eqref{eq:optimality} and \eqref{eq:feasibility}, $P_{T}^\star$ is the optimal average sum rate, $\bbM$ is the size of the maximum independent set of the network and $\eta$ is the dual step size.
\end{proposition}

%%%%%%%%%%%%%%%%%%%%%%%%%%%%%%%%%%%%%%%%%%%%%%%%%%%%%%%%%%%%%%%%%%%%%%%%%%%%%%%%
%%%   P   R   O   O   F   %%%%%%%%%%%%%%%%%%%%%%%%%%%%%%%%%%%%%%%%%%%%%%%%%%%%%%
%%%%%%%%%%%%%%%%%%%%%%%%%%%%%%%%%%%%%%%%%%%%%%%%%%%%%%%%%%%%%%%%%%%%%%%%%%%%%%%%

\begin{proof}
See Appendix \ref{sec:proposition}.
\end{proof}

%\blue{Remove $\lam(T)$ from these bouds.}

% \blue{We may replace this with finite $T$ bounds. Whether we do this or not, we also need to be careful with $P_T$. The convergence result that I know requires comparison to a fixed optimization problem. In fact, I believe that it requires comparison to $P_\infty$. Maybe not an issue since  $P_\infty > P_T$? In any case, result has to checked. It is not a straightforward application of \cite{navidsa,ribeiro2010ergodic} the way it is written.}

%%%%%%%%%%%%%%%%%%%%%%%%%%%%%%%%%%%%%%%%%%%%%%%%%%%%%%%%%%%%%%%%%%%%%%%%%%%%%%%%
%%%   M   A   I   N       M   A   T   T   E   R   %%%%%%%%%%%%%%%%%%%%%%%%%%%%%%
%%%%%%%%%%%%%%%%%%%%%%%%%%%%%%%%%%%%%%%%%%%%%%%%%%%%%%%%%%%%%%%%%%%%%%%%%%%%%%%%

Proposition \ref{prop:limits} establishes that for large $T$, the average rates of the sequence $\bbs^\ddagger(1: T)$ are near-optimal for \eqref{eq:wls}. Asymptotic feasibility follows from Equation \eqref{eq:feasibility} because the multipliers $\bblam(T)$ remain bounded (see Appendix \ref{sec:dualbound}), ensuring the violation term vanishes as $T\rightarrow\infty$. 
%Provided that we consider sufficiently large $T$, Proposition \ref{prop:limits} implies that the average rate achieved by the sequence $\bbs^\ddagger(1: T)$ is near-optimal in \eqref{eq:wls}. As per Equation \eqref{eq:feasibility}, the schedules are feasible because we know $\bblam$ approaches $\bblam^\star$, and therefore $\bblam(T)$ is not growing with $T$ (see Appendix \ref{sec:dualbound}). 
It follows from Proposition \ref{prop:limits} that dual gradient descent finds an optimal solution of \eqref{eq:wls} despite the fact that it does \emph{not} converge to a solution of \eqref{eq:wls}. More to the point, after running dual gradient descent for $T$ iterations we \emph{cannot} claim optimality of the schedule $\bbs^\ddagger(T)$ but we can claim optimality of the schedule sequence $\bbs^\ddagger(1:T)$.

Proposition \ref{prop:limits} admits an alternative interpretation: We can use dual gradient descent as a scheduling algorithm. I.e., we equate dual iteration indexes $u$ with scheduling indexes $t$, so that at time $t=u$ we schedule links as dictated by $\bbs(t) = \bbs^{\ddagger}(u)$. %We then start with an arbitrary initial dual variable $\bblam(1)$ and schedule links as dictated by the Lagrangian maximizing schedule $\bbs^{\ddagger}(1)$ \eqref{eqn_dga_maximizers} at scheduling slot $t=0$. We then update the dual variable as per \eqref{eq:lambdaupdate}. This yields a multiplier $\bblam(1)$ that we use to schedule links as dictated by the Lagrangian maximizing schedule $\bbs^{\ddagger}(1)$ \eqref{eq:lambdaupdate} at scheduling slot $t=1$. 
At a generic scheduling slot $t$ we have a dual variable $\bblam(t) = \bblam(u)$ that dictates the use of the schedule $\bbs^{\ddagger}(t)=\bbs^{\ddagger}(u)$ as determined by the corresponding Lagrangian maximizers in \eqref{eqn_dga_maximizers}. We then update multipliers as per \eqref{eq:lambdaupdate} and proceed to the next scheduling slot $t+1 = u+1$. 

The challenge of using dual gradient descent as a scheduling algorithm is the cost of computing Lagrangian maximizers. The state augmented algorithm of Section \ref{sec_state_augmented_learning} circumvents this problem with the use of a learning parameterization that we train to approximate Lagrangian maximizers.  

%%%%%%%%%%%%%%%%%%%%%%%%%%%%%%%%%%%%%%%%%%%%%%%%%%%%%%%%%%%%%%%%%%%%%%%%%%%%%%%%
%%%   S   E   C   T   I   O   N   %%%%%%%%%%%%%%%%%%%%%%%%%%%%%%%%%%%%%%%%%%%%%%
%%%%%%%%%%%%%%%%%%%%%%%%%%%%%%%%%%%%%%%%%%%%%%%%%%%%%%%%%%%%%%%%%%%%%%%%%%%%%%%%

\section{State Augmented Learning}\label{sec_state_augmented_learning}
Iterations between computing Lagrangian maximizers $\bbs^{\ddagger}(u)$ \eqref{eqn_dga_maximizers} and updating the dual variable $\bblam$ \eqref{eq:lambdaupdate} shape an algorithm that solves \eqref{eq:wls}. The focus shifts towards how to efficiently compute the Lagrangian maximizers for any $\bblam$. To this end, introduce a parametric function class $\bbPhi(\bbA, \bblam; \bbH)$ so that resource allocations $\bbs$ are functions of the interference matrix $\bbA$ and the multiplier $\bblam$ as determined by the choice of parameter $\bbH$
\begin{align}\label{eqn_learning_parameterization_abstract}
    \bbs = \bbPhi(\bbA, \bblam; \bbH).
\end{align}

Given the graph structure intrinsic to our problem, we use a Graph Neural Network (GNN) as the architecture (see Appendix \ref{sec:gnns}). The model will receive the network configuration via the adjacency matrix $\bbA$ and an input graph signal $\mathbf{x}\in\mathbb{R}^K$ supported on the nodes. GNNs are deterministic models, i.e. if the graph signal $\mathbf{x}$ remains constant and we assume the network structure does not change, the output of the model will be the same for all times. This requires explicitly incorporating time-varying components to the parameterization. 

A clever way to make the model's output stochastic is by replacing the input signal $\mathbf{x}$ with the dual variable $\bblam$, a technique known as state augmentation \cite{calvofullana2023stateaugmentedconstrainedreinforcement, navidsa}. At each iteration $u$ the GNN will process each link weighed by the violation of its individual constraint. The obtained scheduling decisions will adapt over iterations to attend undersatisfied links. Note that stable training would be prevented if $\bm\lambda$ is not considered an input, as the abrupt variations of the dual variables induce equally abrupt changes the scheduling decisions.
The incorporated \textit{augmented} parameterization is used as a proxy for $\bbs^\ddagger(u)$. The instantaneous Lagrangian is adjusted as follows:

\begin{align}\label{eqn_time_t_lagrangian_augmented}
    \ccalM \big( \, \bbPhi(\bbA, \bblam; \bbH), \bblam \, \big) 
        & = (\bbone + \bblam)^\top  \Big[\,\bm\Phi(\bbA, \bblam;\bbH) \nonumber \\
        &\quad \odot[\mathbf{1}-\bbA\bm\Phi(\bbA, \bblam;\bbH)]_+  \, \Big],
\end{align}
where the term associated with $\bbDelta$ is omitted as it does not affect the maximization. 

Our proposed learning approach involves an offline training phase and an online execution phase. During the training phase we adopt the instantaneous Lagrangian $\ccalM(\bbs, \bblam)$ as a loss to measure the fit of the resource allocation $\bbs = \bbPhi(\bbA, \bblam; \bbH)$. We then consider a distribution of network realizations $p_{\bbA}$ and a distribution of Lagrange multipliers $p_{\bblam}$ and search for the parameter $\bbH^\star$ that minimizes the average of the instantaneous Lagrangian loss over the choice of network and multiplier,
\begin{align}\label{eqn_offline_learning}
    \bbH^\star  = \argmax_{\bbH} \mbE_{\bbA, \bblam} 
                      \Big[ \, \ccalM \big( \, \bbPhi(\bbA, \bblam; \bbH), \bblam \, \big) \, \Big].
\end{align}

We begin with parameters $\bbH_1$. At each epoch, we obtain the scheduling decisions via the parameterization for a random $\bblam$. During training, we sample $\bblam$ from an arbitrary probability distribution $p_{\bblam}$. This ensures the model is robust to different levels of constraint violation for different links. The adjusted instantaneous Lagrangian can be computed via \eqref{eqn_time_t_lagrangian_augmented}. Stochastic gradient ascent solves the maximization of \eqref{eqn_offline_learning}, updating the parameters in the direction of steepest ascent with step size $\zeta$. The training procedure of the proposed State Augmented GNN (SAGNN) is detailed in Algorithm \ref{alg:train}.

\begin{algorithm}
  \caption{Training Phase for SAGNN Link Scheduling}\label{alg:train}
  \begin{algorithmic}[1]
    \State{\textbf{Input:} Number of training iterations $N$, primal learning rate $\zeta$. 
    }
    \State{\textbf{Initialize:} $\bbH(1)$ randomly}
      \For{$n=1,\cdots, N$}
      \State{Randomly sample $\bblam\sim p_{\bblam}$.}
      % \For{$t=0,\cdots,T-1$}
      \State{Generate scheduling decisions $\bm\Phi(\bbA, \bbH(n);\bblam)$.}
      % \EndFor
      \State{Use Eq. \eqref{eqn_time_t_lagrangian_augmented} to compute the instantaneous Lagrangian $\ccalM(\bbH(n), \bblam)$.}
    \State {Update the primal parameters:
        $$ \bbH(n+1) = \bbH(n) + \zeta\nabla_{\bbH} \ccalM(\bbH(n), \bblam).$$}
      \EndFor
      \State {\textbf{Return:} Optimal model parameters $\bbH^\star=\mathbf{H}(N).$}
  \end{algorithmic}
\end{algorithm}
\begin{algorithm}
  \caption{Execution Phase for SAGNN Link Scheduling}\label{alg:test}
  \begin{algorithmic}[1]
    \State{{\textbf{Input:} Optimal model parameters ${\bbH}^\star$, number of times $T$, dual learning rate $\eta$}}
    \State{Initialize: $\bblam(1) = \textbf{0}$}
      \For{$t=1,\cdots,T$ }
      % \State{$m = \lfloor t / T_0\rfloor$}
      \State{Sample the model and store $\bbs(t) = \bm\Phi(\bbA, \bbH^\star;\bblam(t))$.}
        \State{Update the dual variables via Eq. \eqref{eqn_online_dual_update}:}
        \begin{align*}%\label{eq:lambdaupdateaugmented}
    \bblam(t+1) = \bigg[\, \bblam(t)  - \eta \Big[ \bbs(t) \odot \big[ \mathbf{1} - \bbA\bbs(t) \big]_+ \! - \bbDelta \Big] \, \bigg]_+.
\end{align*}
      \EndFor 
      \State \textbf{Return:} Sequence of scheduling decisions $\bbs(1:T)$.
  \end{algorithmic}
\end{algorithm}

During the online execution phase we consider a time varying Lagrange multiplier $\bblam(t)$ and schedule links as dictated by the optimal parameter $\bbH^\star$,
\begin{align}\label{eqn_online_scheudling}
    \bbs(t) = \bbPhi(\bbA, \bblam; \bbH^\star).
\end{align}
After executing this schedule we proceed to update the dual variables according to the constraint slacks associated with the link schedule $\bbs(t)$,
\begin{align}\label{eqn_online_dual_update}
    \bblam(t+1) = \bigg[\, \bblam(t)  - \eta \Big[ \bbs(t) \odot \big[ \mathbf{1} - \bbA\bbs(t) \big]_+ \! - \bbDelta \Big] \, \bigg]_+.
\end{align}
We initialize $\bblam(1) = \mathbf{0}$. For each iteration $u$ (which is equated to scheduling indexes $t$), use the model trained with \eqref{eqn_online_scheudling} to obtain the corresponding scheduling decisions $\bbs(1)=\bbPhi(\bbA,  \bblam(1); \bbH^\star)$. This enables the update of the dual variable via \eqref{eqn_online_dual_update}. The graph signal $\bblam(u)$ is the input of the model in the following iteration $u+1$. The execution procedure of SAGNN is summarized in Algorithm \ref{alg:test}. 

We say that \eqref{eqn_offline_learning}-\eqref{eqn_online_dual_update} is a state augmented learning algorithm because in addition to scheduling channels as a function of the network $\bbA$, we also schedule channels as a function of the multiplier $\bblam(t)$. We can therefore think of $\bblam(t)$ as an internal state that augments the network realization $\bbA$ to guide scheduling decisions.

The results in Section \ref{sec_dual_gradient_descent} explain why state augmented learning [cf.  \eqref{eqn_offline_learning}-\eqref{eqn_online_dual_update}] solves \ref{eq:wls}. According to the results in Section \ref{sec_dual_gradient_descent}, we obtain schedules that are $\ccalO(1/\eta T)$-feasible and $\ccalO(1/\eta T)$-optimal if we implement the Lagrangian maximizing schedules in \eqref{eqn_dga_maximizers} and the multiplier updates as in \eqref{eq:lambdaupdate}. %State augmented learning determines schedules $\bbs = \bbPhi(\bbA, \bblam; \bbH)$ that maximize the instantaneous Lagrangian over a distribution of network and multiplier realizations [cf. \eqref{eqn_offline_learning}] and updates dual variables according to \eqref{eqn_online_dual_update}. The dual variable updates in \eqref{eq:lambdaupdate} and \eqref{eqn_online_dual_update} are analogous and, if the parametric function class is sufficiently expressive, schedules $\bbs(t) = \bbPhi(\bbA, \bblam(t); \bbH^*)$ and $\bbs^\ddagger(u)$ are close. Thus, state augmented learning is an approximate implementation of dual gradient descent and it follows from Proposition \ref{prop:limits} that it results in scheduling decisions that are $\ccalO(1/\nu T)$-feasible and $\ccalO(1/\nu T)$-optimal.
State augmented learning [cf.  \eqref{eqn_offline_learning}-\eqref{eqn_online_dual_update}] is an odd choice, but one that is justified because more straightforward alternatives have drawbacks. The two following remarks explain this. 

%%%%%%%%%%%%%%%%%%%%%%%%%%%%%%%%%%%%%%%%%%%%%%%%%%%%%%%%%%%%%%%%%%%%%%%%%%%%%%%%
%%%   R   E   M   A   R   K   %%%%%%%%%%%%%%%%%%%%%%%%%%%%%%%%%%%%%%%%%%%%%%%%%%
%%%%%%%%%%%%%%%%%%%%%%%%%%%%%%%%%%%%%%%%%%%%%%%%%%%%%%%%%%%%%%%%%%%%%%%%%%%%%%%%

\begin{remark} \label{rmk_parameterization_unworkable} \normalfont

The most natural choice is to choose schedules according to a learning parameterization without state augmentation, i.e., we choose schedules as  $\bbs(t) = \bbPhi_1(\bbA; \bbH)$. This is unworkable because the schedule $\bbs(t)$ is the same for all times $t$ and schedules that solve \eqref{eq:wls} almost never have this property. Indeed unless the rate requirement is $\Delta_i = 0$ for some links, satisfying rate constraints \emph{requires} alternating channel access. 

\end{remark}

%%%%%%%%%%%%%%%%%%%%%%%%%%%%%%%%%%%%%%%%%%%%%%%%%%%%%%%%%%%%%%%%%%%%%%%%%%%%%%%%
%%%   R   E   M   A   R   K   %%%%%%%%%%%%%%%%%%%%%%%%%%%%%%%%%%%%%%%%%%%%%%%%%%
%%%%%%%%%%%%%%%%%%%%%%%%%%%%%%%%%%%%%%%%%%%%%%%%%%%%%%%%%%%%%%%%%%%%%%%%%%%%%%%%

\begin{remark} \label{rmk_parameterization_expensive} \normalfont

An alternative choice of learning parameterization is $\bbs(1:T) = \bbPhi_2(\bbA; \bbH)$. I.e., we train a parametric function classes to find entire schedules $\bbs(1:T)$ instead of individual scheduling decisions $\bbs(t) = \bbPhi_1(\bbA; \bbH)$ -- which as we explain in Remark \ref{rmk_parameterization_unworkable} is unworkable. This is workable but expensive to train because we are trying to learn a function class whose output is of dimension $KT$. This is in contrast to the outputs of the state augmented learning parameterization in \eqref{eqn_learning_parameterization_abstract} which are of dimension $K$.

\end{remark}

%  \section{Transferability Properties}
% \label{sec:transf}
% \input{transferability}

\section{Numerical Experiments}
\label{sec:simu}

We conduct experiments to evaluate the proposed algorithm.\footnote{Code is available at \url{https://github.com/romm32/SAGNN}.} We generated a dataset of Random Geometric Graphs (RGGs), a standard model for real-world wireless networks where two nodes are connected if and only if they are within a fixed communication radius of each other. Our implementation places nodes on a regular grid, with Gaussian noise added to the positions to introduce irregularity, and forms edges according to the distances between nodes. This construction provides a balance between controllability and scalability in network complexity, as we can decide the variance of the added noise. For each communication graph, we compute its corresponding conflict graph. The distribution of node degrees in the conflict graph reflects the complexity of the scheduling task, as it governs interference. Each graph in the training and testing datasets has an average of $K\simeq500$ links. We leave the study of transferability across scales for future work, as discussed in Section \ref{sec:conclusion}. 

Unless otherwise stated, we assign an identical minimum transmission requirement to all links, i.e., $\bm\Delta = \Delta \mathbf{1}_K$. The choice of the transmission requirement 
$\Delta$ must be made carefully. If set too high, the problem may become infeasible. Conversely, if it is too low, it will have little to no effect on the resulting scheduling policy. To guide the selection of $\Delta$, we consider the average degree of the conflict graph as an indicator of the network's interference level, which loosely bounds the feasible scheduling capacity \cite{jafar2013topological}. %Additionally, we use the fraction of links in the graph’s Maximum Independent Set (MIS) as an upper bound for the achievable transmission requirement.

We provide a comprehensive evaluation of our algorithm’s performance, testing on graphs with Gaussian degree distributions. This introduces higher structural variability and increases scheduling complexity. We first analyze the training process to confirm that optimization proceeds as expected and converges successfully (Section \ref{sec:hyp1-trainingconverges}). To assess the feasibility of the obtained solutions, we evaluate the trained model on unseen data and observe that constraints are satisfied for the vast majority of links. The remaining violations are minor (Section \ref{sec:hyp2-constraintsatisfaction}). 

Since we imitate optimal solutions, it is sensical to compare our method against existing heuristic approaches (Section \ref{sec:hyp3-baselines}). One of the differences between our approach and other baselines is that, as explained in Section \ref{sec_state_augmented_learning}, we do not learn to solve the problem directly. We instead learn to compute Lagrangian maximizers. Nonetheless, the maximization does not involve the transmission requirement $\bbDelta$ (see Section \ref{sec_dual_gradient_descent}). 

We study the generalization of our policy under different transmission requirements (Section \ref{sec:hyp4-differentdeltas}). Meeting the requirement while avoiding interference implies the policy switches across different subsets of links as time advances. We characterize the behavior at the start and end of the considered time window to observe the performance over time (Section \ref{sec:hyp5-timeanalysis}). Furthermore, we find that we schedule avoiding collisions the majority of the time (Section \ref{sec:hyp6-collisions}). The frequency of collisions can be further reduced with a simple masking technique.

\subsection{Training}
\label{sec:hyp1-trainingconverges}
The model is trained using Algorithm \ref{alg:train} for $N=100$ epochs. A GNN is used as the main architecture, followed by a sigmoid activation function to produce values in the continuous range $[0, 1]$ (see Appendix \ref{sec:implementationdetails} for more details on the architecture and the implementation). During the first epoch, dual variables $\bm\lambda$ are sampled from a uniform distribution $\mathcal{U}[0, 2]^K$. In subsequent epochs, we instead sample from empirical dual variable vectors. We carry out the evaluation (Algorithm \ref{alg:test}) at every epoch to monitor the performance of the model. This gives access to plausible dual variables we can feedback into our training. As these empirical dual variables were obtained via Equation \eqref{eq:lambdaupdate}, the value of $\Delta$ is implicitly reintroduced into training. Appendix~\ref{sec:impactlambda} presents an analysis of how the distribution of dual variables affects performance across graphs with different levels of regularity. The parameters $\bbH$ are optimized via primal gradient ascent with $\zeta = 5\times10^{-5}$ using the Adam optimizer \cite{adam}.

Evaluation is carried out considering $T=200$ in Algorithm \ref{alg:test}. The training requires continuous scheduling decisions to allow for backpropagation in the neural network. However, a threshold of 0.5 is used at evaluation time to binarize which links are scheduled for transmission, allowing for more accurate performance measurement. The minimum transmission requirement is $\Delta=\{0.1, 0.125, 0.15\}$ for different trained models, where the task complexity increases with $\Delta$. Unless otherwise stated, we use the model with $\Delta=0.1$ for subsequent experiments. The value of $\Delta$ is needed to compute the subgradient with Equation \eqref{eqn_subgradient}. Dual variable updates use a step size of $\eta=2$. To mitigate the effects that a few links with infeasible constraints might have on the overall solution, we consider a resilient formulation of the problem \cite{resilience,gnnsforrrm2}. The result is the addition of an L1 regularization term to the instantaneous Lagrangian in Equation \eqref{eqn_time_t_lagrangian_augmented}. As the term is independent of the parameters $\bbH$, the training algorithm remains unchanged. The updates for the dual variable are as follows:
\begin{align}\label{eq:lambdaupdateaugmentedresilience}
    \bblam(u+1) 
        &= \Big[\, \bblam(u) - \eta[  \, \bbg(\bbH^\star, \bblam(u))+\bblam\alpha] \,\Big]_+  . 
\end{align}
The resilience factor $\alpha$ is fixed at 0.05 for all transmission requirements. For more details on the resilient formulation, see Appendix \ref{sec:implementationdetails}.

Figure \ref{fig:constraints} shows constraint violation results for models trained for different minimum transmission requirements, evaluated on a validation dataset at the end of each epoch. More specifically, we present the percentage of links for which $\mathbf{\overline{r}} < \mathbf{\Delta}$. The curves represent the mean and standard deviation over three independent experiments with different random seeds. Higher values of $\Delta$ increase problem complexity and approach infeasibility. As training progresses, the number of constraint violations decreases, indicating that the model effectively learns to schedule links over time successfully.

\begin{figure}
    \centering
\includegraphics[width=0.9\linewidth]{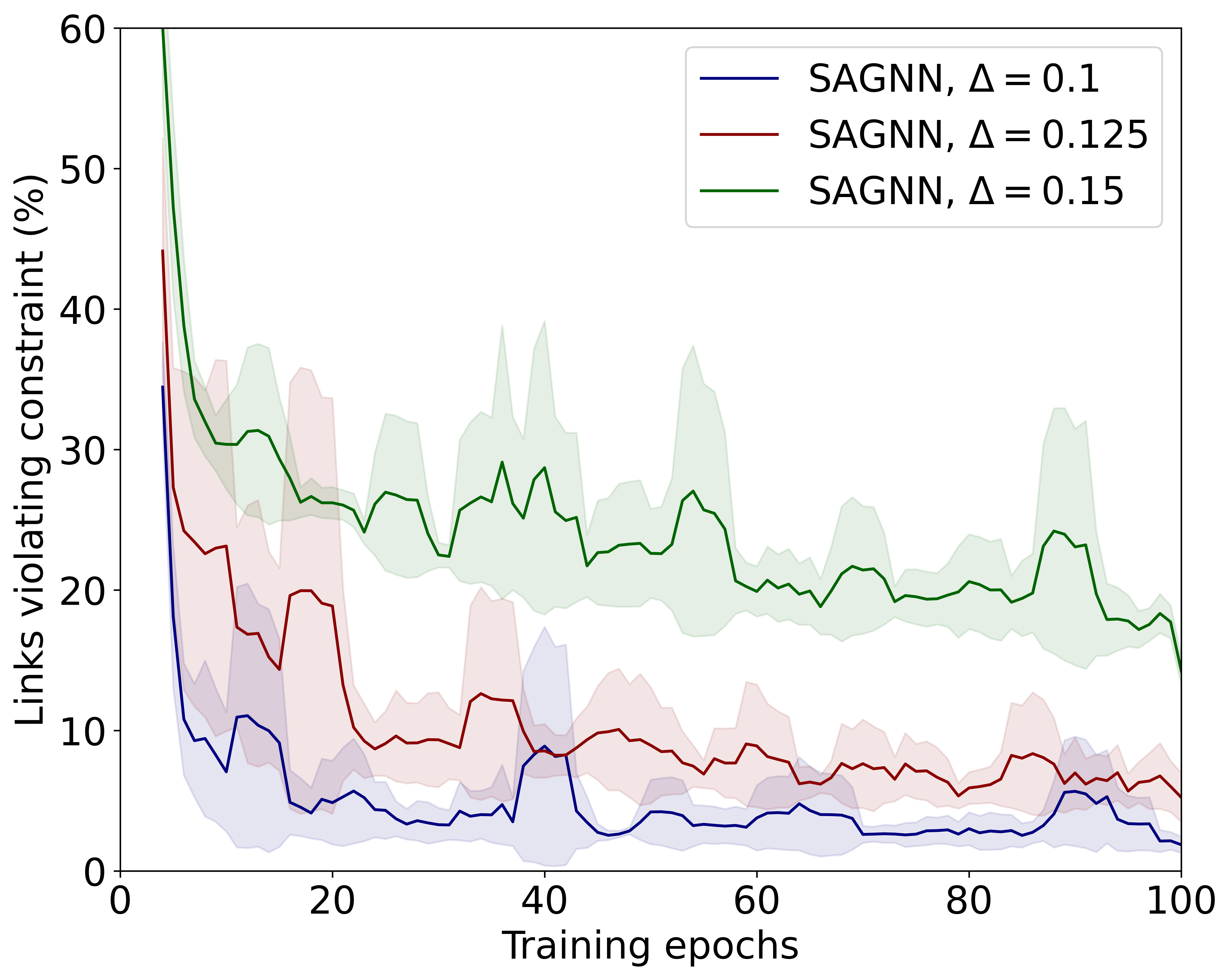}
    \caption{Evolution of constraint violation during training for different minimum transmission requirements. The mean across three experiments is presented, with the standard deviation shown in a lighter color. A running average of factor 5 was applied on the curves to enhance clarity.}
    \label{fig:constraints}
\end{figure}

\begin{figure}
    \centering
    \includegraphics[width=0.9\linewidth]{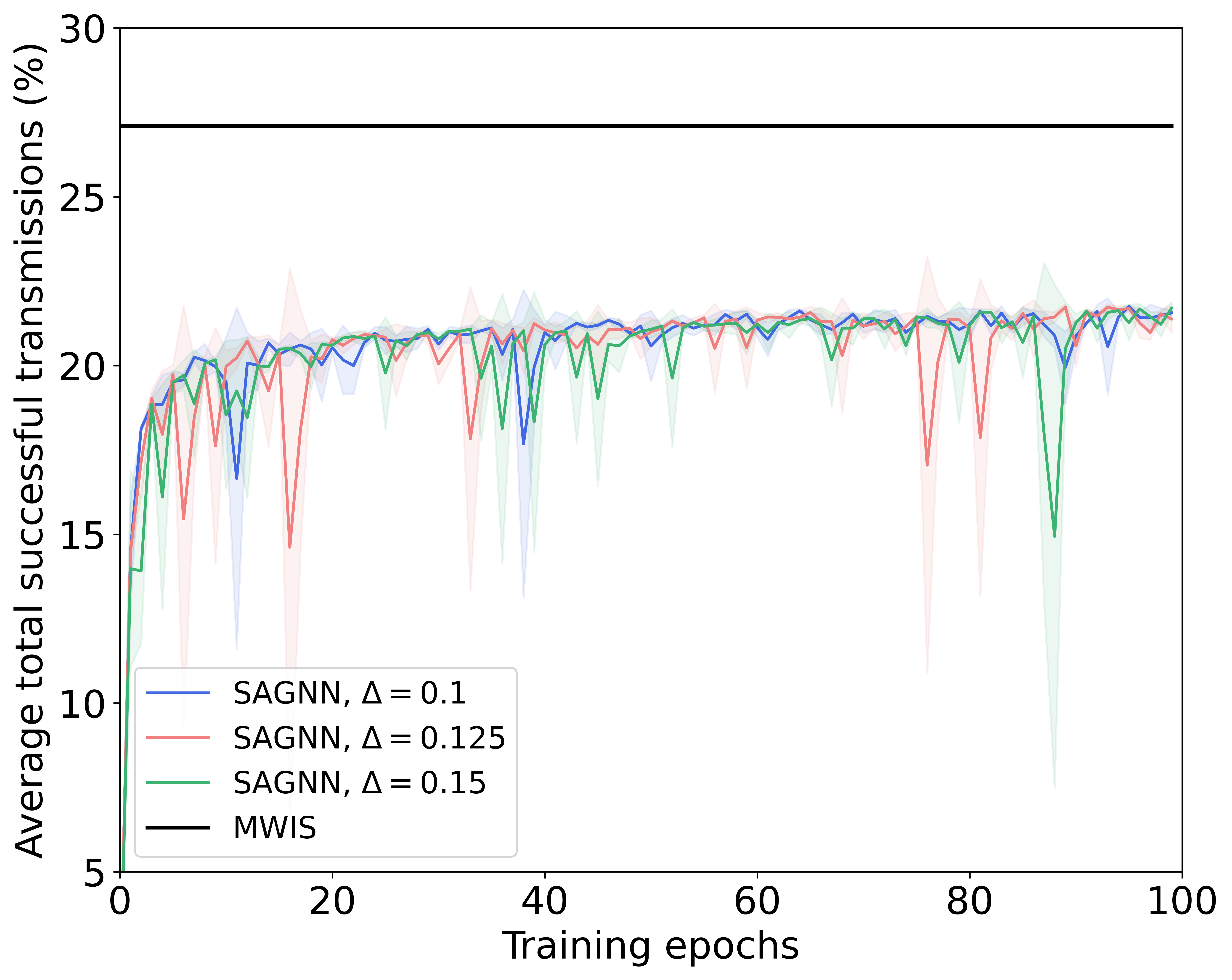}
    \caption{Evolution of the average successful transmissions in the network during training. The mean across three experiments is presented, with the standard deviation shown in a lighter color. SAGNN shows convergence close to the average MWIS in the dataset.}
    \label{fig:of}
\end{figure}

For each model, Figure \ref{fig:of} shows the achieved average rates $\mathbf{1}^\top\overline{\mathbf{r}}$. Our proposed method, SAGNN, converges to a value close to the average Maximum Weighted Independent Set (MWIS) size of the graphs used, which serves as an upper bound on the achievable objective value. The comparison is further developed in Section \ref{sec:hyp3-baselines}. 

\subsection{Constraint Satisfaction}
\label{sec:hyp2-constraintsatisfaction}
\begin{figure}
    \centering
    \includegraphics[width=0.98\linewidth]{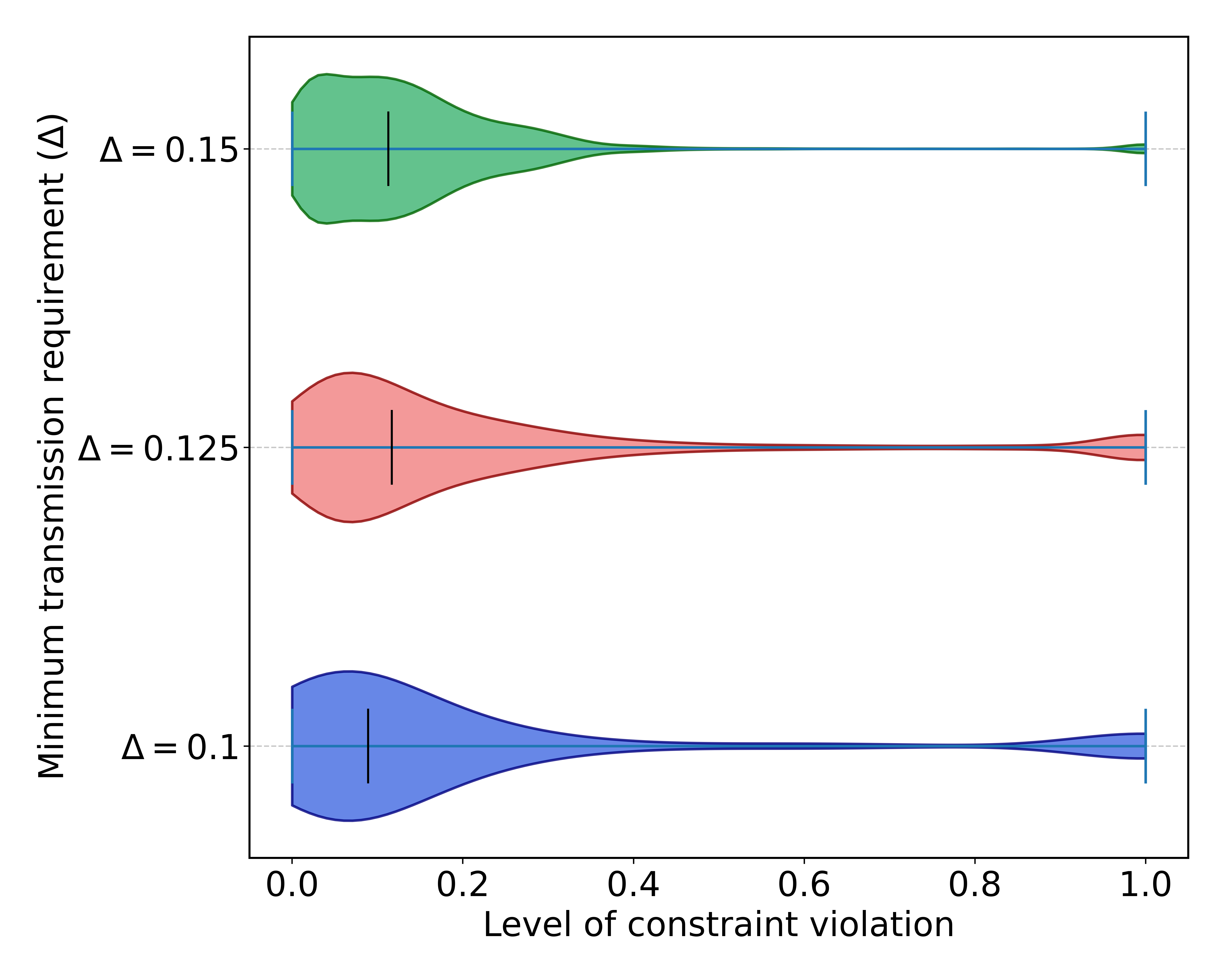}
    \caption{Box plot of the distribution of values for links violating the constraint. The values are scaled such that 1 is the largest possible value (which would indicate a complete violation of amount $\Delta$).}
    \label{fig:boxplot}
\end{figure}

\begin{figure}
    \centering
    \includegraphics[width=0.9\linewidth]{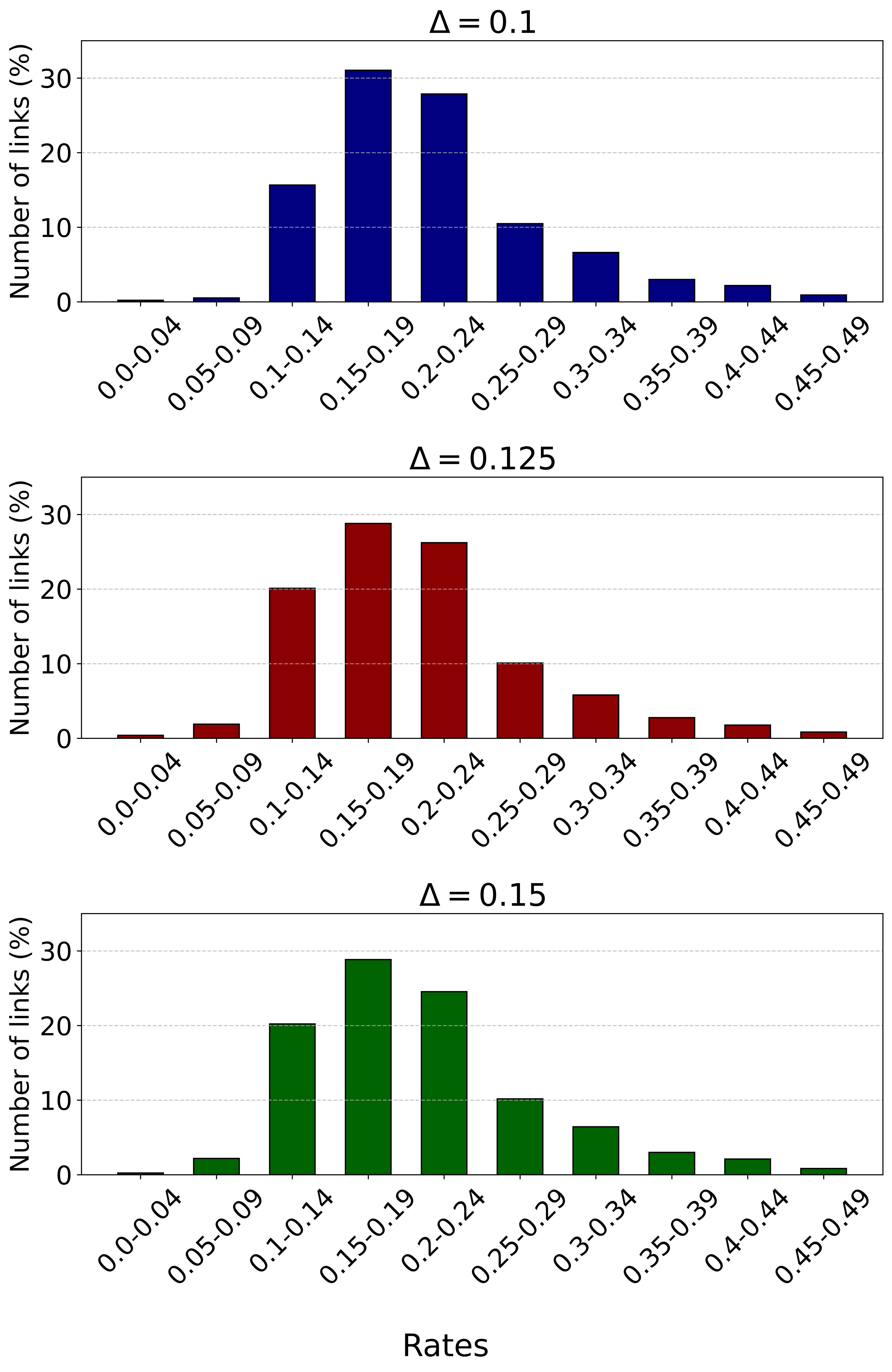}
    \caption{Successful transmission rates achieved by links for varying requirements.}
    \label{fig:rates}
\end{figure}

The curves shown in Figure \ref{fig:constraints} indicate that a small fraction of links still fails to meet the minimum transmission requirement at the end of training. To assess the severity of the remaining constraint violations, we analyze the violation magnitude $\mathbf{\Delta} - \mathbf{\overline{r}}$. Figure \ref{fig:boxplot} shows the distribution of constraint violation levels. For a more consistent comparison, violations are normalized such that a value of 1 represents a complete failure to transmit (i.e., a shortfall equal to $\Delta$). Across all transmission requirements, the vast majority of violations remain below $15\%$, with only a few outliers. These results provide strong evidence that the proposed algorithm effectively learns to satisfy the transmission constraints. Furthermore, the introduction of resilience ensures that links at risk of violating the constraint still receive high transmission rates, rather than being disregarded by the model due to infeasibility. This is illustrated in Figure \ref{fig:rates}.

\subsection{Benchmark Comparison}
\label{sec:hyp3-baselines}
We compare the proposed algorithm against several heuristic baselines, listed below. We do not include any learning-based approach, as the state-of-the-art baseline FPLinQ is used for comparison in the literature. 

\begin{enumerate} [i.]
    \item \textit{p}-persistent scheduling: Determine a probability of transmission for each link, correlated to its degree of conflict. Sample Bernoulli trials according to these probabilities to schedule transmissions.
    \item \textit{p}-persistent with collision avoidance (CA): Determine the scheduling using the \textit{p}-persistent strategy. For every pair of conflicting links, randomly turn one of them off. 
    \item FPlinQ \cite{shen2017fplinq}: State-of-the-art baseline for link scheduling based on fractional programming. It maximizes the instantaneous sum rate. 
    \item Maximum weight independent set (MWIS) scheduling \cite{mwis1, mwis2, mwis3}: Optimal solution to the problem where the links are weighted by their corresponding entry as the dual variable $\bblam$. %To implement it, we initialize all links with a fixed weight, then find and schedule the MWIS for the first time step. Given these scheduling decisions, the weights are updated decreasing them for successfully scheduled links and increasing them for the remaining links. We repeat this procedure for all time steps to achieve good time-averaged performance.  
\end{enumerate}

The performance of our algorithm and the baselines are summarized in Table \ref{tab:baselines}, averaging results over $T=200$ time steps for 10 graphs from the test set (due to the high computational cost of the baseline methods). Although SAGNN does not attain the highest average sum rate, it remains competitive while offering inference runtimes that are orders of magnitude faster. Beyond its strong performance, SAGNN is also naturally suited for decentralized deployment since it relies only on local information \cite{zwang2021decentralized}. In contrast, decentralizing FPLinQ or MWIS would require each node to replicate complex centralized computations, which could lead to inconsistent network views and degraded scheduling decisions. Additionally, SAGNN operates without requiring full channel state information (CSI), unlike FPLinQ, and thus incurs significantly lower communication and information overhead.

\begin{table}[t]
    \centering
    \begin{tabular}{lccc}
      \toprule
           & \begin{tabular}[c]{@{}c@{}}Constraint \\ violation (\%)\end{tabular} 
             & \begin{tabular}[c]{@{}c@{}}Objective \\ function (\%)\end{tabular} 
             & \begin{tabular}[c]{@{}c@{}}Runtime\end{tabular} \\ 
      \midrule
      $p$-persistent     & 97.57$\pm$0.62  & 2.69$\pm$0.08   & $0.04\pm 4.88$ $\mu$s   \\
      $p$-persistent CA  & 85.47$\pm$1.44  & 6.27$\pm$0.14   & $0.04\pm 3.73$ $\mu$s   \\
      FPLinQ             & 74.91$\pm$0.75  & 25.09$\pm$0.75  & $15.3 \pm 0.50$ ms              \\
      MWIS               & 1.55$\pm$0.34   & \textbf{27.12$\pm$0.21} & $14.22\pm2.30$ s          \\
      SAGNN              & \textbf{0.75$\pm$0.45} & 21.91$\pm$0.26 & $1.11 \pm 0.03$ ms        \\
      \bottomrule
    \end{tabular}
    \caption{Performance evaluation averaged across 10 graphs from the testing dataset, showing constraint violation and objective function in percentage of links. We also present the average runtime required for generating a link scheduling for one time step, for one graph.}
    \label{tab:baselines}
\end{table}

\subsection{Robustness to Different Link Priorities}
\label{sec:hyp4-differentdeltas}
\begin{figure}
    \centering
    \includegraphics[width=\linewidth]{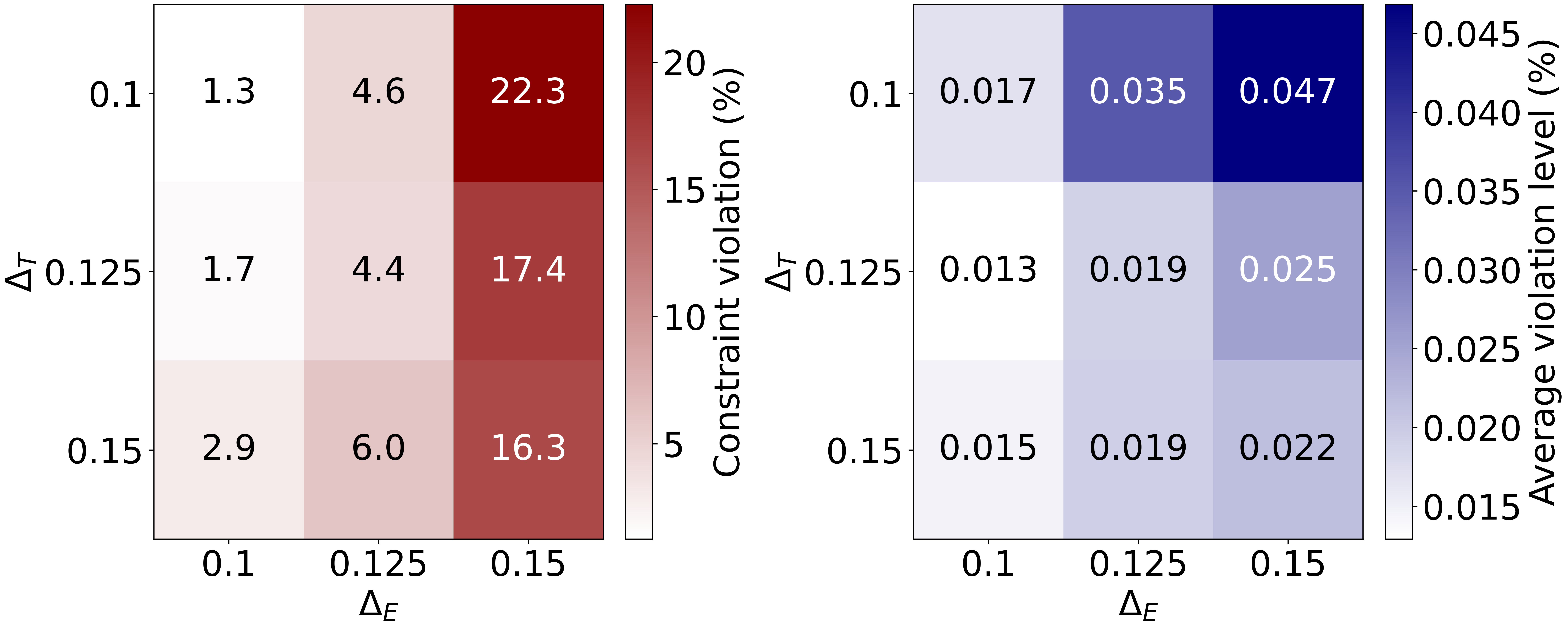}
    \caption{In red, percentage of constraint violation when a model with $\Delta_{T(rain)}$ is evaluated in different $\Delta_{E(xecution)}$. In blue, the average of the violation level for each corresponding case of $\Delta_T, \Delta_E$. Performance is best for the diagonal, where the requirement aligns for training and execution.}
    \label{fig:constrainttransf}
\end{figure}

Figure \ref{fig:constrainttransf} demonstrates the performance of the proposed method when the requirement for training $\Delta_T$ differs from the requirement of execution $\Delta_E$. We present the percentage of links for which $\mathbf{\overline{r}} < \mathbf{\Delta}$, and further quantify the severity of these violations. As expected, performance degrades when the evaluation constraint deviates from the training value. Nevertheless, all three models demonstrate strong generalization, maintaining low average violation levels even when evaluated under mismatched transmission requirements.

\begin{figure}
    \centering
    \includegraphics[width=0.95\linewidth]{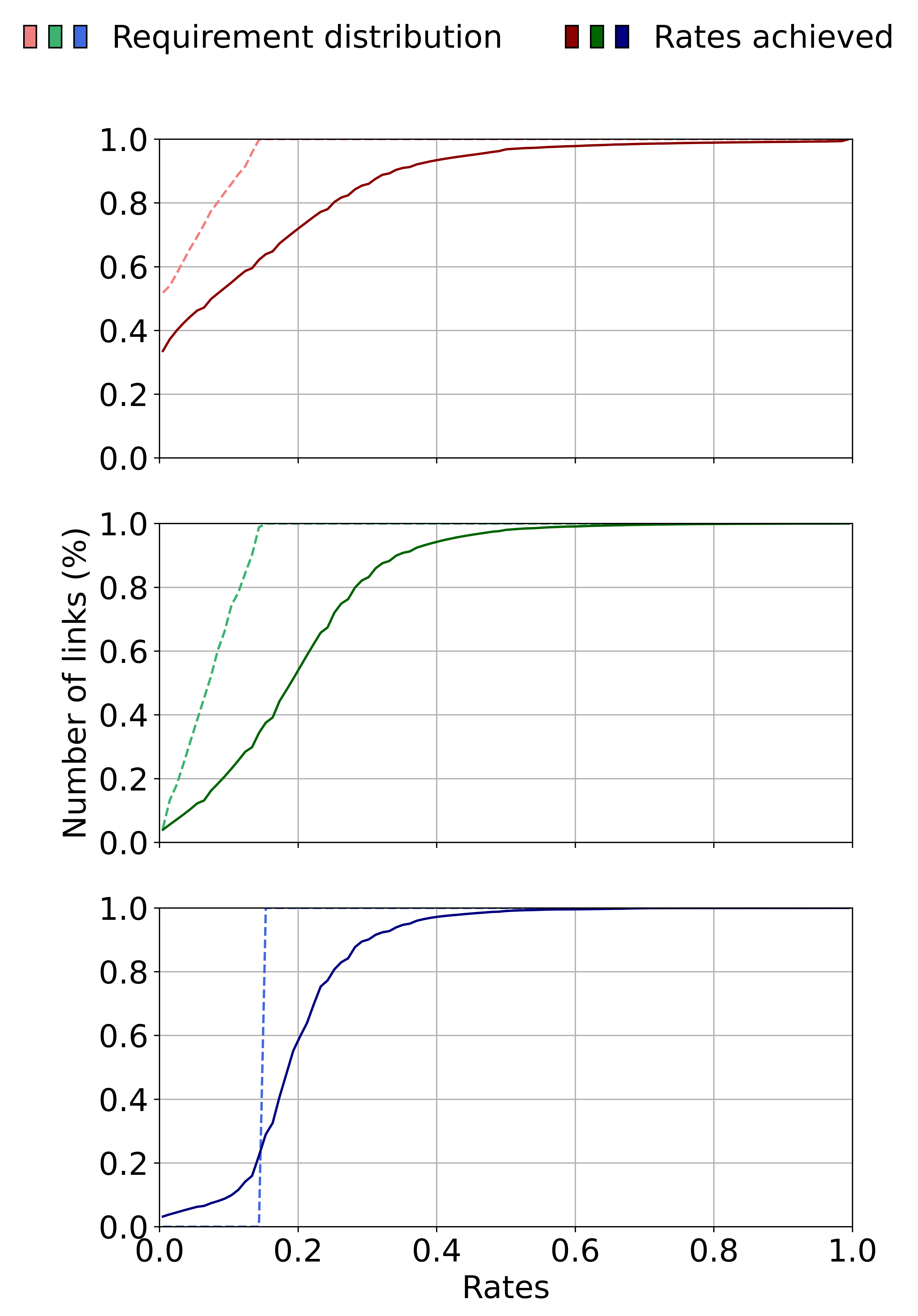}
    \caption{Robustness of the model to different minimum transmission requirements across links. In lighter colors, the corresponding distribution of requirements. In darker colors, the percentages of links that achieved different transmission rates. In the top plot (red), half the links have $\Delta=0$, while the other half have a requirement uniformly distributed in $[0, 0.15]$. In the middle plot (green), all links have a requirement sampled from a uniform distribution in $[0, 0.15]$. In the bottom plot (blue), we show the results for all links having $\Delta=0.15$. }
    \label{fig:otherdeltas}
\end{figure}

Previous evaluations have considered that the minimum transmission requirement meets $\bbDelta=\textbf{1}\Delta$, i.e., all links are subject to the same constraint. We test the robustness of the trained model under more realistic conditions where links are assigned different transmission priorities. Firstly, we randomly select half of the links and assign them zero requirement—effectively removing their scheduling priority—while the remaining links receive requirement values sampled uniformly between 0 and 0.15, consistent with the feasible range observed in previous experiments. In the second configuration, all links are assigned requirement values from this same uniform distribution. Figure \ref{fig:otherdeltas} displays the cumulative density function (CDF) of the requirements in light colors with a dashed line and the rates achieved in darker colors. The results show that the model adapts its scheduling behavior to align with the assigned priorities, effectively generalizing to these new problem instances. The achieved CDF is almost always smaller than the required CDF, which implies most of the links are achieving higher rates than assigned. 

\subsection{Properties of Scheduling Policies}
\label{sec:hyp5-timeanalysis}
Previous results have demonstrated the success of our algorithm in terms of performance averaged over time, as required by the problem formulation in Equation \eqref{eq:wls}. It is reasonable to also study the behavior of the algorithm in the time domain.

\begin{figure}
    \centering
    \includegraphics[width=0.95\linewidth]{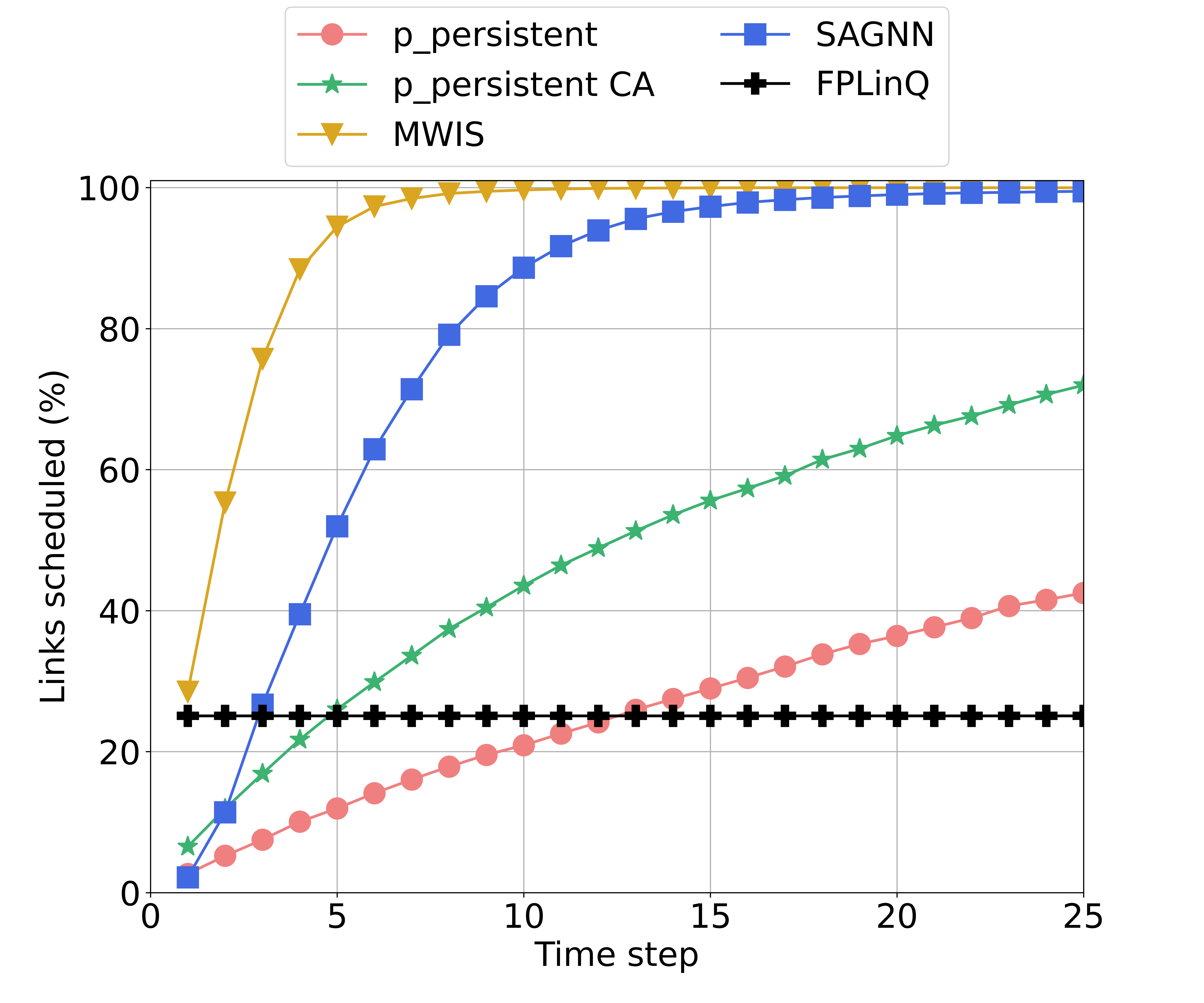}
    \caption{Percentage of links scheduled at least once at the first $T=25$ steps of evaluation in a subset of 10 graphs from the test dataset.
    }
    \label{fig:baselines}
\end{figure}

Figure \ref{fig:baselines} illustrates the evolution of the percentage of links scheduled at least once during the initial evaluation times. SAGNN rapidly approaches $100\%$, matching the performance of MWIS. Unlike FPLinQ, which repeatedly schedules the same subset of links and thus limits diversity in channel access, SAGNN distributes channel access more evenly across links. 

Figure \ref{fig:illustration} illustrates a subset of nodes and edges in an example conflict graph. Scheduled links for the last three time steps of evaluation are highlighted with a thicker black line. The obtained policy is indeed achieving a switching behavior between different independent sets in the conflict graph. Moreover, the links scheduled are associated to high values of the dual variable, shown in shades of red. Observe that after a link has been scheduled, its corresponding multiplier decreases in the next time step.

\subsection{Interference Evaluation}
\label{sec:hyp6-collisions}
\begin{figure}
    \centering
    \includegraphics[width=0.9\linewidth]{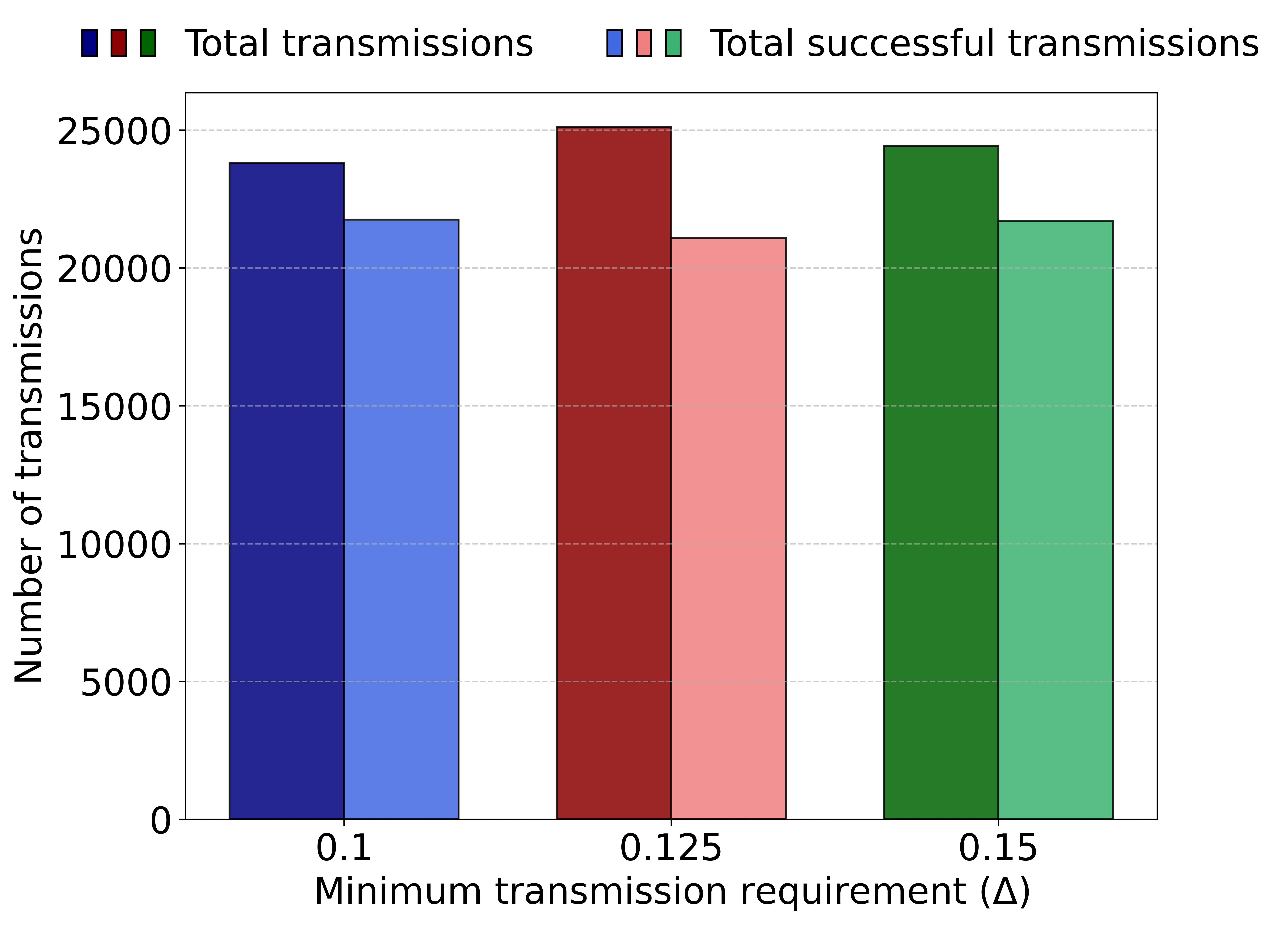}
    \caption{Number of total and successful transmissions for different requirements. The average across 100 graphs is presented for different values of $\Delta$.}
    \label{fig:plot8-succvstotal}
\end{figure}

Figure \ref{fig:plot8-succvstotal} compares the number of successful transmissions to the total number of transmission attempts. The evaluation is performed on the test dataset using 100 unseen graphs, with transmission counts averaged across these samples. Results show that the success ratio remains stable as the minimum transmission requirement $\Delta$ increases. This demonstrates that the algorithm effectively avoids scheduling conflicting links, minimizing wasted channel resources and sustaining a high rate of successful transmissions. 

Further gains in performance can be achieved by masking out interfering transmissions. If the adjacency matrix $\mathbf{A}$ of the conflict graph is available, we can compute the conflict term $\mathbf{A}\mathbf{s}_{\text{binary}}(t)$ to identify collisions at time step $t$. For each set of colliding links, we deactivate all but the one involved in less collisions, ensuring that only non-conflicting links are scheduled. This guarantees that all active transmissions are successful, reducing wasted power and increasing the overall achieved rates. The masking procedure is simple, can be implemented in a decentralized fashion, and enhances the model’s effectiveness. Table \ref{tab:masked} demonstrates the performance improvement, where results for our method are repeated from Table \ref{tab:baselines} for comparison. Evaluation is conducted on the same 10 unseen graphs. A drawback of this approach is the increased average runtime required to resolve collisions and finalize the schedule for each time step $t$.

\begin{table}[]
    \centering
    \begin{tabular}{lccc}
      \toprule
           & \shortstack{Constraint \\ violation (\%)} 
           & \shortstack{Objective \\ function (\%)} 
           & \shortstack{Runtime} \\
      \midrule
      SAGNN        & 0.75$\pm$0.45   & 21.91$\pm$0.26   & \textbf{1.11 $\pm$ 0.03 ms} \\
      SAGNN masked & \textbf{0.16$\pm$0.25} & \textbf{23.11$\pm$0.23} & $4.15 \pm 10.8$ ms \\
      \bottomrule
    \end{tabular}
    \caption{Performance results averaged over 10 unseen graphs for SAGNN and a masked version of the algorithm. We present mean and standard deviation.}
    \label{tab:masked}
\end{table}

\section{Conclusions}
\label{sec:conclusion}
We presented a formulation for the link scheduling problem that optimizes a long-term performance metric subject to per-link rate constraints. More importantly, we introduced an algorithm that leverages dual dynamics to obtain sequences of schedules that solve the problem. Our theoretical analysis demonstrates that the average rate achieved by our scheduling policy is asymptotically feasible and near-optimal as the time horizon $T$ increases. These findings are corroborated by numerical experiments, which show that our policy effectively maximizes the network sum rate while satisfying minimum transmission requirements on average. We observe that constraint violations are rare and, when they occur, remain minimal in magnitude. Compared to heuristic approaches, our algorithm achieves superior constraint satisfaction and reduced runtime while maintaining a competitive average sum rate. Furthermore, the proposed method demonstrates strong generalization across varying transmission requirements and yields high effective rates.

%corroborate our theoretical findings with numerical results. The proposed algorithm converges successfully (Section \ref{sec:hyp2-constraintsatisfaction}). From this starting point, we show that several hypothesis hold. (i) Constraint violations are rare (Section \ref{sec:hyp2-constraintsatisfaction}). For under-scheduled links, the level of constraint violation is minor. (ii) Performance is improved when compared to existing heuristics (Section \ref{sec:hyp3-baselines}). Our algorithm achieves better constraint satisfaction and faster runtime, all the while maintaining strong average sum rate performance. (iii) The policy generalizes to different transmission requirements (Section \ref{sec:hyp4-differentdeltas}). (iv) Our policy adapts over time (Section \ref{sec:hyp5-timeanalysis}) and (v) achieves good effective rates (Section \ref{sec:hyp6-collisions}). We corroborate that collisions are rare in our solutions and their frequency can be further reduced by applying a simple masking technique.

The link scheduling problem considered in this paper is very relevant to the topological interference management and index coding problems, which have been extensively studied in the information theory literature~\cite{jafar2013topological,naderializadeh2014interference,yi2015topological,yang2017topological,yi2018topological,geng2021multilevel,shan2025revisiting}. The implications of this work for better approximations and characterization of the (generalized) degrees of freedom (DoF) and capacity regions of multi-user interference channels~\cite{etkin2008gaussian,naderializadeh2014itlinq,geng2015optimality} are interesting research directions that we leave for future work.

Our algorithm employs a State-Augmented Graph Neural Network (SAGNN). Beyond stability and scalability, a critical advantage of GNNs in this context is transferability—the ability to generalize to larger or varying network topologies. However, existing theoretical analyses of GNN transferability focus primarily on dense graphs \cite{ruiz2021transferability, levie2019transferability}, a regime that does not accurately model the sparse, geometric nature of wireless networks. While this transferability has been extensively observed in experiments \cite{eisen2020optimal, eisen20transferability, transf, wang22}, establishing its theoretical foundations in wireless scenarios remains an open challenge that recent works have begun to address \cite{camargo2025graphneuralnetworkslarge}.
%%%%%%%%%%%%%%%%%%%%%%%%%%%%%%%%%%%%%%%%%%%%%%%%%%%%%%%%%%%%

\urlstyle{same}
\bibliographystyle{IEEEtran}
\bibliography{references}

\appendix
{\section{Appendix}
 \subsection{Proofs}
\label{sec:proposition}
We present the proofs of the propositions from Section \ref{sec:lagrangiandualdomain}.
\begin{pproposition}\label{prop_lagrangian_maximizersappendix}
Let $\bbs^\dagger(1:T, \bblam)$ be a maximizing \eqref{eq:maxofL} schedule of the Lagrangian \eqref{eq:lag} and let $\bbs^\dagger(t, \bblam)$ be the time $t$ entry of this schedule. Time $t$ entries can be computed by maximizing the instantaneous Lagrangian \eqref{eqn_time_t_lagrangian},  
\begin{align}\label{eq:maxofLtimeapp}
    \bbs^\dagger(t, \bblam) 
        ~\equiv~ \argmax_{ \bbs \in \{0,1\}^K } 
                     \ccalM (\bbs, \bblam ).
\end{align}
Consequently, the set of time $t$ and time $v$ Lagrangian maximizers contain the same schedules. I.e.,
\begin{align}\label{eq:maxofLtime-tuapp2}
    \bbs^\dagger(t, \bblam) ~\equiv~ \bbs^\dagger(v, \bblam),
\end{align}
for all dual variables $\bblam$ and all time pairs $t$ and $v$.

\end{pproposition}

\begin{proof}
From Equation \eqref{eq:lag} we have:
\begin{align}
    \ccalL \Big(\, \bbs(1: T), \bblam \, \Big) 
        = \frac{1}{T} \sum_{t=1}^{T} 
              \ccalM(\bbs (t), \bblam).
\end{align}
Observe that to obtain $\bbs^\dagger(v, \bm{\lambda})$ we maximize the Lagrangian $\ccalL (\, \bbs(1: T), \bblam \,)$ evaluated at time $v$:
\begin{align}
    \bbs^\dagger(v, \bblam) 
        =& \argmax_{\bbs(v)\in\{0,1\}^K} 
              \ccalL \Big(\, \bbs(v),\, \bblam \, \Big)\nonumber\\
        =& \argmax_{\bbs(v)\in\{0,1\}^K} 
               \sum_{t=v}^{v} 
              \ccalM(\bbs (t), \bblam)\nonumber\\
        =& \argmax_{\bbs(v)\in\{0,1\}^K}  
              \ccalM(\bbs (v), \bblam).
\end{align}

This can be repeated for all times $v\in\{1, \dots, T\}$. Nonetheless, from our definition of the instantaneous Lagrangian from Equation \eqref{eqn_time_t_lagrangian} we have that it does not exhibit time dependence. The set of time $t$ and time $v$ Lagrangian maximizers contain the same schedules for all times $t, v$ and dual variable $\bm{\lambda}$:
\begin{align}\label{eq:maxofLtime-tuapp}
    \bbs^\dagger(t, \bblam) ~\equiv~ \bbs^\dagger(v, \bblam).
\end{align}
\end{proof} 

\begin{pproposition}\label{prop_subgradientapp}

The instantaneous constraint slack associated with the instantaneous Lagrangian maximizer $\bbs^\ddagger(u)$ \eqref{eqn_dga_maximizers} is a subgradient of the dual function $d(\bblam)$ at $\bblam = \bblam(u)$. I.e., the vector 
\begin{align}\label{eqn_subgradientapp}
    \bbg(\bblam(u)) ~=~   \rofs{\bbs^\ddagger(u)} ~-~ \bbDelta
\end{align}
is such that $\bbg^\top(\bblam(u))(\bblam(u) - \bblam^\star) \geq d(\bblam(u)) - d(\bblam^\star) \geq 0$.

\end{pproposition}

\begin{proof}
    Let us first consider the term $\bbg^\top(\bblam(u))(\bblam(u) - \bblam^\star)$:

    \begin{align} 
    \bbg^\top(\bblam(u))(\bblam(u) - \bblam^\star)
        &= (\rofs{\bbs^\ddagger(u)} - \bbDelta)^\top \nonumber\\
        &\qquad(\bblam(u) - \bblam^\star).\label{one}
    \end{align}

    We defined $d(\bblam(u))=\ccalL(\bbs^\dagger(1:T, \bblam(u)), \bblam(u))$:
    \begin{align}
        d(\bblam(u)) - d(\bblam^\star) =& \frac{1}{T} \sum_{t=1}^{T} 
              \ccalM(\bbs^\dagger (t), \bblam(u)) \nonumber\\
              &\quad- \frac{1}{T} \sum_{t=1}^{T} 
              \ccalM(\bbs^\dagger (t), \bblam^\star)\nonumber\\
        =& \frac{1}{T} \sum_{t=1}^{T} (\bblam(u) - \bblam^\star)^\top\nonumber\\
        &\qquad \label{two}(\bbs^\dagger(t)\odot[\mathbf{1}-\bbA\bbs^\dagger(t)]_+ - \bbDelta).
    \end{align}
    Observe that both \eqref{one} and \eqref{two} share the factor $\bblam(u)-\bblam^\star$, which does not depend on $t$. It remains to see that
    \begin{align}\label{eq:inequalitiesprop2}
        \bbs^\ddagger(u)\odot[\mathbf{1}-\bbA\bbs^\ddagger(u)]_+  \geq \frac{1}{T} \sum_{t=1}^{T} \bbs^\dagger(t)\odot[\mathbf{1}-\bbA\bbs^\dagger(t)]_+ \geq \bbDelta 
    \end{align}
    Note that we are assuming constraints are not violated, i.e. the achieved average rates are higher than the transmission requirement $\bbDelta$. Moreover, the first inequality in Equation \eqref{eq:inequalitiesprop2} arises from the definitions of $\bbs^\ddagger$ and $\bbs^\dagger$ and observing the average sum rate will be smaller or equal to the optimal rate for a specific time or iteration index.
\end{proof}

\begin{pproposition}
    Let $\bar\bbr[\bbs(1: T)]$ be the average rates in \eqref{eqn_trajectory_rates} obtained by executing the sequence of Lagrangian maximizing schedules $\bbs^\ddagger(\bblam(u))$. These maximizers are associated with the sequence of multipliers $\bblam(1: T)$ obtained from running the subgradient updates in \eqref{eq:lambdaupdate}. The average rates $\bar\bbr[\bbs(1: T)]$ are asymptotically nearly optimal
\begin{align}\label{eq:optimalityapp}
\bbone^\top\bar\bbr\big[\bbs^\ddagger(1: T)\big]
            \geq P^\star_T - \frac{\eta}{2}\|\bbM-\bbDelta\|^2-\frac{|\bblam(1)|^2}{2\eta T},
\end{align}
and asymptotically feasible, 
\begin{align}\label{eq:feasibilityapp}
        \bar\bbr\big[\bbs^\ddagger(1: T)\big]
            \geq \mathbf{\Delta}- %\frac{\bblam(T)-\bblam(1)}{\eta T}.
            \frac{\bblam(T)}{\eta T}.
\end{align}
In \eqref{eq:optimalityapp} and \eqref{eq:feasibilityapp}, $P_{T}^\star$ is the optimal average sum rate, $\bbM$ is the size of the maximum independent set of the network and $\eta$ is the dual step size.
\end{pproposition}

\begin{proof}
    We first prove optimality as per Equation \eqref{eq:optimalityapp}. To show optimality it suffices to bound the optimality gap. Consider Lagrangian maximizers $\bbs^\ddagger(1: T)$, which maximize the instantaneous Lagrangian, and write the following in terms of the complete Lagrangian.
    \begin{align}
        \frac{1}{T} \sum_{t=1}^{T} \bbone^\top&\rofs{\bbs^\ddagger(t)}\nonumber\\
        +& \bblam^\top(t)\big[ \rofs{\bbs^\ddagger(t)}-\bbDelta\big] \nonumber\\ 
        & \geq \frac{1}{T} \sum_{t=1}^{T} \bbone^\top\rofs{\bbs(t)}\nonumber\\
        &\quad+ \bblam^\top(t)\big[ \rofs{\bbs(t)}-\bbDelta\big].
    \end{align}
    The second term on the RHS is non-negative if we assume $\bbs(1:T)$ feasible. Reordering and applying Equation \eqref{eqn_subgradientapp} for the second term on the LHS:
    \begin{align}
        \frac{1}{T} \sum_{t=1}^{T} &{\bbs^\ddagger(t)[\bbone-\bbA\bbs^\ddagger(t)]_+} \geq \nonumber\\
        \frac{1}{T} &\sum_{t=1}^{T} {\bbs(t)[\bbone-\bbA\bbs(t)]_+} -\bblam^\top(t)\bbg(\bblam(t)).
    \end{align}
    Evaluate $\bbs(1:T)=\bbs^\star(1:T)$ the optimal schedules:
    \begin{align}
        \frac{1}{T} \sum_{t=1}^{T} &{\bbs^\ddagger(t)[\bbone-\bbA\bbs^\ddagger(t)]_+} \geq 
        P_T^\star-\frac{1}{T}\sum_{t=1}^{T}\bblam^\top(t)\bbg(\bblam(t)).
    \end{align}
    To show that the optimality gap is bounded, we simply need to bound the RHS of
    \begin{align}\label{eq:chosenone}
        P^\star_T - \bbone^\top\bar{\bbr}\big[\bbs^\ddagger(1: T)\big]
            \leq \frac{1}{T}\sum_{t=1}^{T}\bblam^\top(t)\bbg(\bblam(t)).
    \end{align}
    We begin by bounding the norm of $\bblam(t+1)$.
    \begin{align}\label{eq:lambda_recursion}
        \|\bblam(t+1)\|^2 =& \| [\bblam(t)-\eta\bbg(\bblam(t))]_+\|^2\nonumber\\
        \leq& \|\bblam(t)\|^2 + \eta^2\|\bbg(\bblam(t))\|^2 \nonumber\\&- 2\eta\bblam^\top(t)\bbg(\bblam(t)),
    \end{align}
    where we expand the square and remove the projection on the positive orthant. We can show the second term is bounded because $\eta$ is finite and $\bbg(\bblam(t))$ is bounded considering a vector whose norm is the size of the maximum independent set of the conflict graph, $\bbM\in \{0, 1\}^K$.
    \begin{align}
    \|\bbg(\bblam(t))\|^2 = \|\rofs{\bbs^\ddagger(t)}-\bbDelta\|^2 \leq \|\bbM-\bbDelta\|^2
    \end{align}

    We rewrite our bound for $\|\bblam(t+1)\|^2$. As it is true for all time indexes, we can unroll the recursion on the RHS of Equation \eqref{eq:lambda_recursion}.
    \begin{align}\label{eq:equationpositive}
        0 \leq \|\bblam(t+1)\|^2 \leq \|\bblam(t)\|^2 &+ \eta^2\|\bbM-\bbDelta\|^2\\&-2\eta\bblam^\top(t)\bbg(\bblam(t)).\nonumber
    \end{align}
    The term on the LHS of \eqref{eq:equationpositive} is positive because it is a squared norm. We sum from $1$ to $T$ and use \eqref{eq:equationpositive} iteratively. Further we divide over $2\eta T$ to obtain the following:
    \begin{align}
         \frac{1}{T}\sum_{t=1}^{T}\bblam^\top(t)\bbg(\bblam(t)) \leq \frac{\|\bblam(1)\|^2}{2\eta T} -& \frac{\|\bblam(T)\|^2}{2\eta T} \nonumber\\&+ \frac{\eta}{2}\|\bbM-\bbDelta\|^2.
    \end{align}
    We return to our previous objective from Equation \eqref{eq:chosenone}.
    \begin{align}\label{eq:finalopt}
        P^\star_T - \bbone^\top\bar{\bbr}\big[\bbs^\ddagger(1: T)\big]
            \leq& \frac{1}{T}\sum_{t=1}^{T}\bblam^\top(t)\bbg(\bblam(t))\nonumber\\
            \leq&\frac{\eta}{2}\|\bbM-\bbDelta\|^2+\frac{\|\bblam(1)\|^2-\|\bblam(T)\|^2}{2\eta T}.
            %=& \mathcal{O}(\frac{1}{{\sqrt{T}}}).
    \end{align}
    The first term on the RHS of \eqref{eq:finalopt} is constant and proportional to $\eta$. The second term vanishes asymptotically as $T\rightarrow\infty$ (it is $\mathcal{O}(\frac{1}{T})$ for a fixed $\eta$, see Proposition \ref{prop_bound}). Consequently, the upper bound is dominated by the first term, yielding an optimality gap of $\mathcal{O}(\eta)$. This recovers \eqref{eq:optimalityapp}.

    We show feasibility starting from the update rule of the dual variable in \eqref{eq:lambdaupdate}.
    \begin{align}
    \bblam(t+1) = &\big[\bblam(t)-\eta(\rofs{\bbs^\ddagger(t)}-\bbDelta) \big]_+\nonumber\\
    \geq& \bblam(t)-\eta(\rofs{\bbs^\ddagger(t)}-\bbDelta), 
    \end{align}
    where the inequality follows by removing the projection onto the positive orthant. Summing this inequality over $t=1, \dots, T$ yields:

    \begin{align}
        \bblam(T) \geq \bblam(1) -\eta\sum_{t=1}^{T} \rofs{\bbs^\ddagger(t)} +\eta T\bbDelta .
    \end{align}
    Rearranging the terms to isolate the accumulated rates, we obtain:
\begin{align}
    \eta \sum_{t=1}^{T} \rofs{\bbs^\ddagger(t)} \geq \eta T \mathbf{\Delta} + \bblam(1) - \bblam(T).
\end{align}
Dividing both sides by $\eta T$ recovers the average rate $\bar\bbr\big[\bbs^\ddagger(1: T)\big]$:
\begin{align}
    \bar\bbr\big[\bbs^\ddagger(1: T)\big] &\geq \mathbf{\Delta} - \frac{\bblam(T)-\bblam(1)}{\eta T}\\
    &\geq \mathbf{\Delta} - \frac{\bblam(T)}{\eta T}.
\end{align}
Since the dual variables $\bblam(T)$ remain bounded within a neighborhood of the optimal multipliers $\bblam^\star$, the second term on the RHS vanishes as $\mathcal{O}(\frac{1}{T})$ for large $T$. This implies that the constraint violation converges to zero, proving that the policy is asymptotically feasible and recovering \eqref{eq:feasibilityapp}.

\end{proof}

\subsection{Bound in the Dual Variables}
\label{sec:dualbound}
We present in Proposition \ref{prop_bound} the bound for the dual variables when updated via Equation \eqref{eq:lambdaupdate} for increasing iterations.

\begin{pproposition}\label{prop_bound}
    The dual variable $\bblam$ strictly approaches the optimal dual variable $\bblam^\star$ via the update in Equation \eqref{eq:lambdaupdate} whenever the current approximation error is sufficiently large. Specifically, the distance to the optimum decreases monotonically,
    \begin{align}
         \|\bblam(u+1)-\bblam^\star\| < \|\bblam(u)-\bblam^\star\|,
    \end{align}
    provided that the current distance satisfies:
    \begin{align}
        \|\bblam(u)-\bblam^\star\| > \frac{\eta\|\bbM-\bbDelta\|^2}{2\varepsilon}.
    \end{align}
    Consequently, the sequence $\{\bblam(u)\}$ does not diverge but oscillates within a bounded neighborhood of $\bblam^\star$.
\end{pproposition}
\begin{proof}
    We analyze the evolution of the dual variables via \eqref{eq:lambdaupdate}.  
    \begin{align}
        \|\bblam(u+1)-\bblam^\star\|^2&= \|[\bblam(u)-\eta\bbg(\bblam(u))]_+-\bblam^\star\|^2\nonumber \\
        &\leq \|\bblam(u)-\eta\bbg(\bblam(u))-\bblam^\star\|^2\nonumber\\
        &=\|\bblam(u)-\bblam^\star\|^2-2\eta\bbg^\top(\bblam(u))(\bblam(u)-\bblam^\star)\nonumber\\
        &\quad+\eta^2\|\bbg(\bblam(u))\|^2.
    \end{align}
    Using Proposition \ref{prop_subgradientapp} and the bound for $\|\bbg(\bblam(u))\|^2$ derived in Appendix \ref{sec:proposition}, we get
    \begin{align}
        \|\bblam(u+1)-\bblam^\star\|^2&\leq\|\bblam(u)-\bblam^\star\|^2-2\eta(d(\bblam(u))-d(\bblam^\star))\nonumber\\
        &\quad+\eta^2\|\bbM-\bbDelta\|^2.
    \end{align}
    Since $d(\bblam(u))-d(\bblam^\star)\geq0$, the distance to the optimum decreases strictly whenever the duality gap is sufficiently large:
    \begin{align}\label{eq:ineq}
        2\eta(d(\bblam(u))-d(\bblam^\star))>\eta^2\|\bbM-\bbDelta\|^2.
    \end{align} 
     Substituting the strict feasibility condition $d(\bblam(u))-d(\bblam^\star)\geq \varepsilon \|\bblam(u)-\bblam^\star\|$ into the inequality in Equation \eqref{eq:ineq}, we observe that the distance to the optimum decreases strictly whenever the current error is sufficiently large:
    \begin{align}
        \|\bblam(u)-\bblam^\star\| > \frac{\eta\|\bbM-\bbDelta\|^2}{2\varepsilon}.
    \end{align}
    This inequality defines a region of guaranteed descent. As long as the dual variable $\bblam(u)$ lies outside this radius, the gradient information dominates the noise introduced by the constant step size, ensuring monotonic convergence toward $\bblam^\star$. Once inside this region, the constant step size prevents further convergence to the exact optimum. Consequently, the sequence $\bblam(u)$ does not diverge but remains uniformly bounded, oscillating within a neighborhood of $\bblam^\star$ with a radius proportional to $\eta$.
\end{proof}

\subsection{Graph Neural Networks}
\label{sec:gnns}
Graph Neural Networks consist of a cascade of layers, each comprising a graph convolutional filter followed by a pointwise nonlinearity \cite{gama2019convolutional,scarselli2008graph}. These filters are polynomials on a matrix representation of the graph, aggregating information from neighboring nodes. The graph shift operator (GSO) \cite{ortega2018graph} serves as this matrix representation, enabling signal diffusion over the graph’s edges. We use the adjacency matrix $\bbA$ of the conflict graph as the GSO. Formally, the graph convolutional filter at the $l$-th layer, with input graph signal $\mathbf{x}_{l-1}$, is expressed as follows:
\begin{align}
    \mathbf{y}_l = \sum_{k=0}^{K-1} h_{lk} \mathbf{A}^k \mathbf{x}_{l-1}, \qquad \mathbf{x}_{l-1}, \mathbf{y}_l \in \mathbb{R}^K,
\end{align}
where $\{h_{lk}\}_{k=0}^{K-1}$ are the learnable parameters of the filter in the $l$-th layer. The output of the graph convolutional filter is then passed through a pointwise nonlinearity $\sigma:\mathbb{R} \rightarrow \mathbb{R}$, resulting in $\mathbf{x}_l = \sigma(\mathbf{y}_l).$

%Given the clear correspondence between a communication network and its graph representation, GNNs are commonly employed to map network scenarios to resource management strategies \cite{wang2022learning, eisen2020optimal, navidsa,shen2020graph}. 

GNNs possess several desirable properties that make them well-suited for real-world applications \cite{gnnprops, gama2019stability, ruiz2021transferability, wang2024geometric, transf}. One such property is scalability, which enables their use in high-dimensional problems. This stems from their ability to exploit the structure encoded in the graph and apply convolutional filters efficiently. GNNs also provide stability, a critical feature for tasks involving large-scale or dynamic systems. A further advantage is transferability: when trained on one graph, a GNN can generalize to different graph structures and signals—provided the model is encapsulated in a filter tensor—while maintaining strong performance. This allows training on one network and deploying on others with similar statistical properties. Perhaps the most fundamental property of GNNs is permutation equivariance, inherited from graph filters. This allows GNNs to operate independently of node labels and to leverage the symmetries present in both graph structures and signals. As a result, they can learn rich representations from a single input, enhancing the model's expressive power.

\subsection{Implementation Details}
\label{sec:implementationdetails}
We adopt a Graph Neural Network (GNN) architecture with $L=3$ layers, each with 256 features. Convolutional stages use TagConv filters \cite{tagconv} of order $3$, followed by Batch Normalization \cite{batchnorm} and a leaky ReLU activation as the pointwise nonlinearity. A sigmoid activation is applied at the output to produce values in the range $[0,1]$. While more complex techniques such as the REINFORCE algorithm and the Gumbel-Softmax \cite{jang2017categoricalreparameterizationgumbelsoftmax} trick were considered, the relaxation to the continuous range proved enough to obtain good performance.

The model is trained for $N=100$ epochs. In each epoch we iterate through the entire training and validation dataset. We generate 300 conflict graphs, and split them equally into train, validation, and test datasets. The average number of links in each network is $K\simeq 500$ links.

\subsection{Dual Variable Sample Distribution}
\label{sec:impactlambda}
In the preliminary experiments presented in \cite{camargo2025wirelesslinkschedulingstateaugmented}, the evaluation was conducted on nearest-neighbor graphs constructed from grid topologies without noise added to positions. This setting significantly simplifies the scheduling problem: nodes exhibit nearly identical interference patterns, and the graph structure is highly regular. Under the assumption that each node in the conflict graph has degree $d$, an effective scheduling policy would consist of identifying $d$ largely disjoint maximum independent sets and alternating among them over time. In such a case, using a uniform distribution for $p_{\bm{\lambda}}$ is reasonable, as all nodes play similar roles. However, as we introduce irregularity by adding noise to the node positions, the homogeneity of the interference patterns breaks down, and uniform scheduling strategies become suboptimal. In these scenarios, the choice of $\bm{\lambda}$ must account for the varying transmission opportunities of different links. While in the experiments in Section \ref{sec:simu} we consider a single dataset with high irregularity, we now consider three datasets with increasing levels of irregularity, induced by different noise levels applied to the grid node positions. The corresponding degree distributions of their conflict graphs are shown in Figure \ref{fig:distributions}.

\begin{figure}[t]
    \centering
    \subfloat[No noise]{%
        \includegraphics[width=0.31\linewidth]{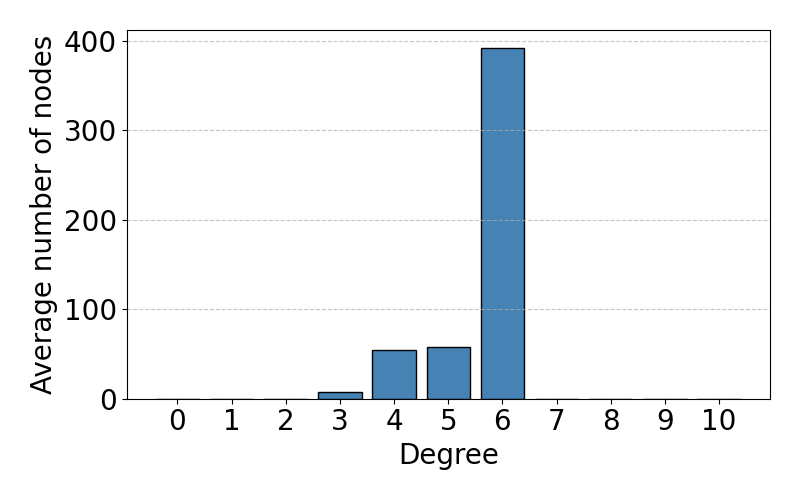}%
        \label{subfig:noise0}
    }
    \hfill
    \subfloat[Moderate noise]{%
        \includegraphics[width=0.31\linewidth]{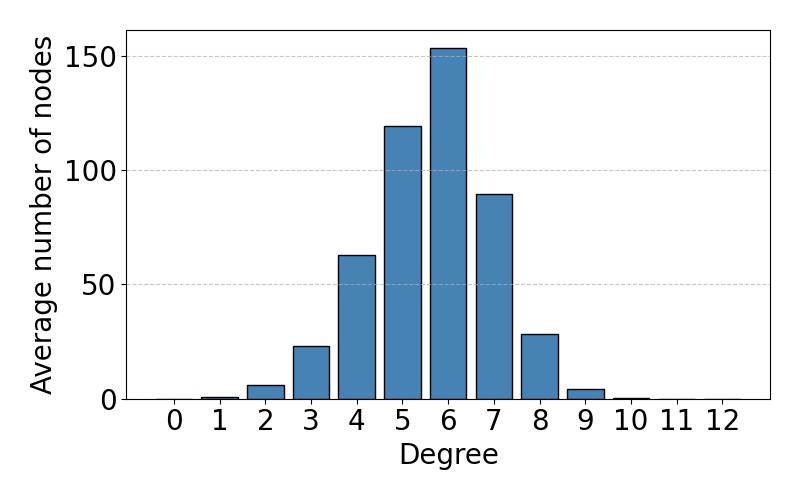}%
        \label{subfig:noise0.0074}
    }
    \hfill
    \subfloat[High noise]{%
        \includegraphics[width=0.31\linewidth]{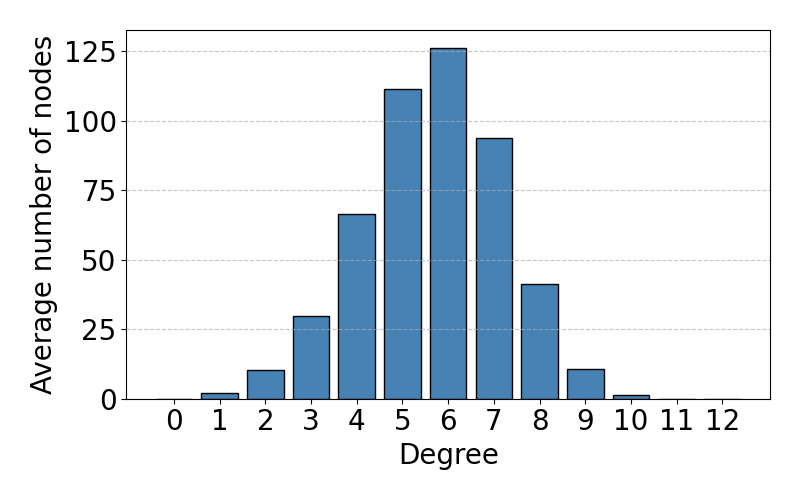}%
        \label{subfig:noise0.01}
    }
    \caption{Degree distributions for conflict graphs with increasing noise in positions.}
    \label{fig:distributions}
\end{figure}

\begin{figure}[t]
    \centering
    \subfloat[No noise]{%
        \includegraphics[width=0.45\linewidth]{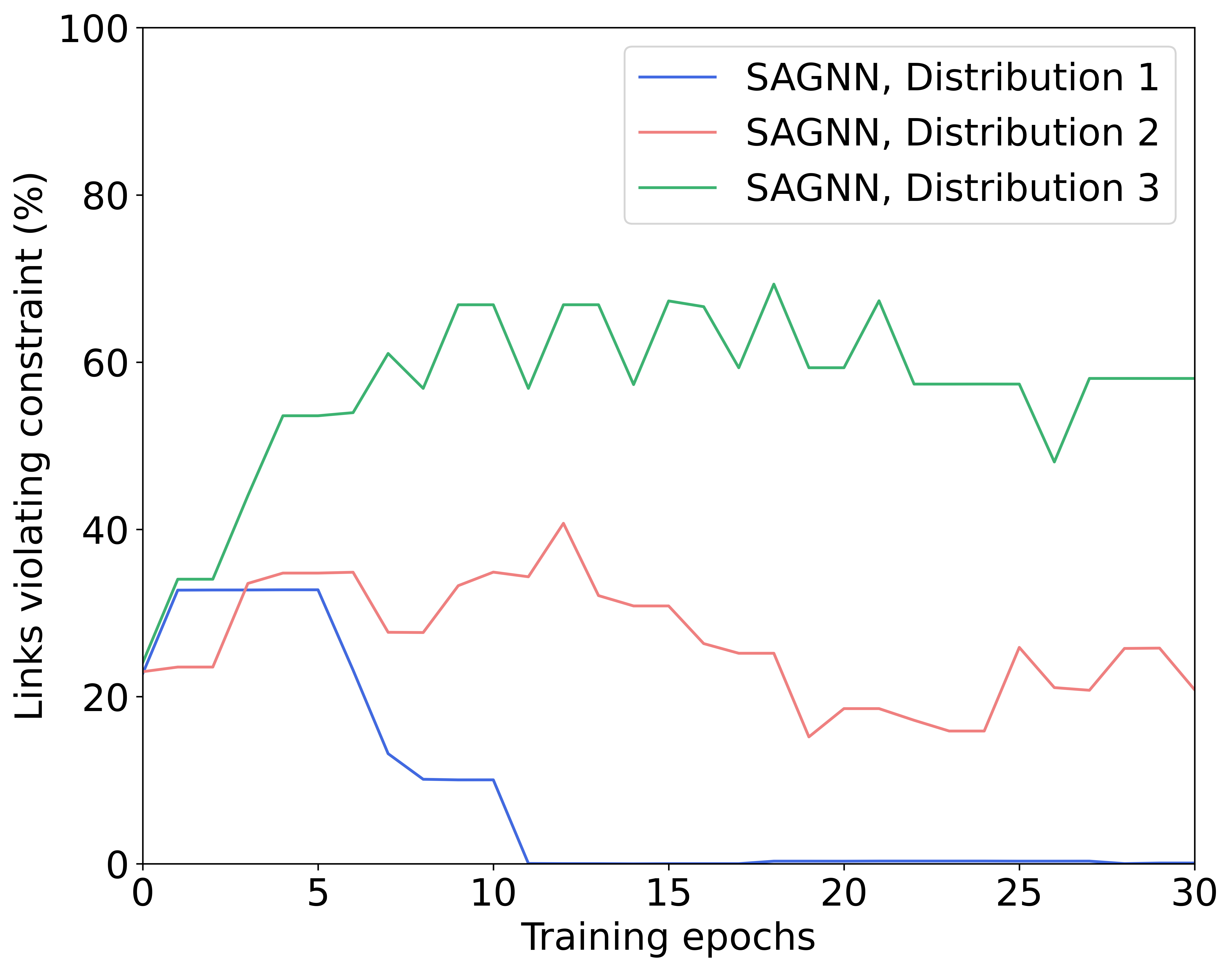}%
        \label{subfig:noise01}
    }
    \hfill
    \subfloat[Moderate noise]{%
        \includegraphics[width=0.45\linewidth]{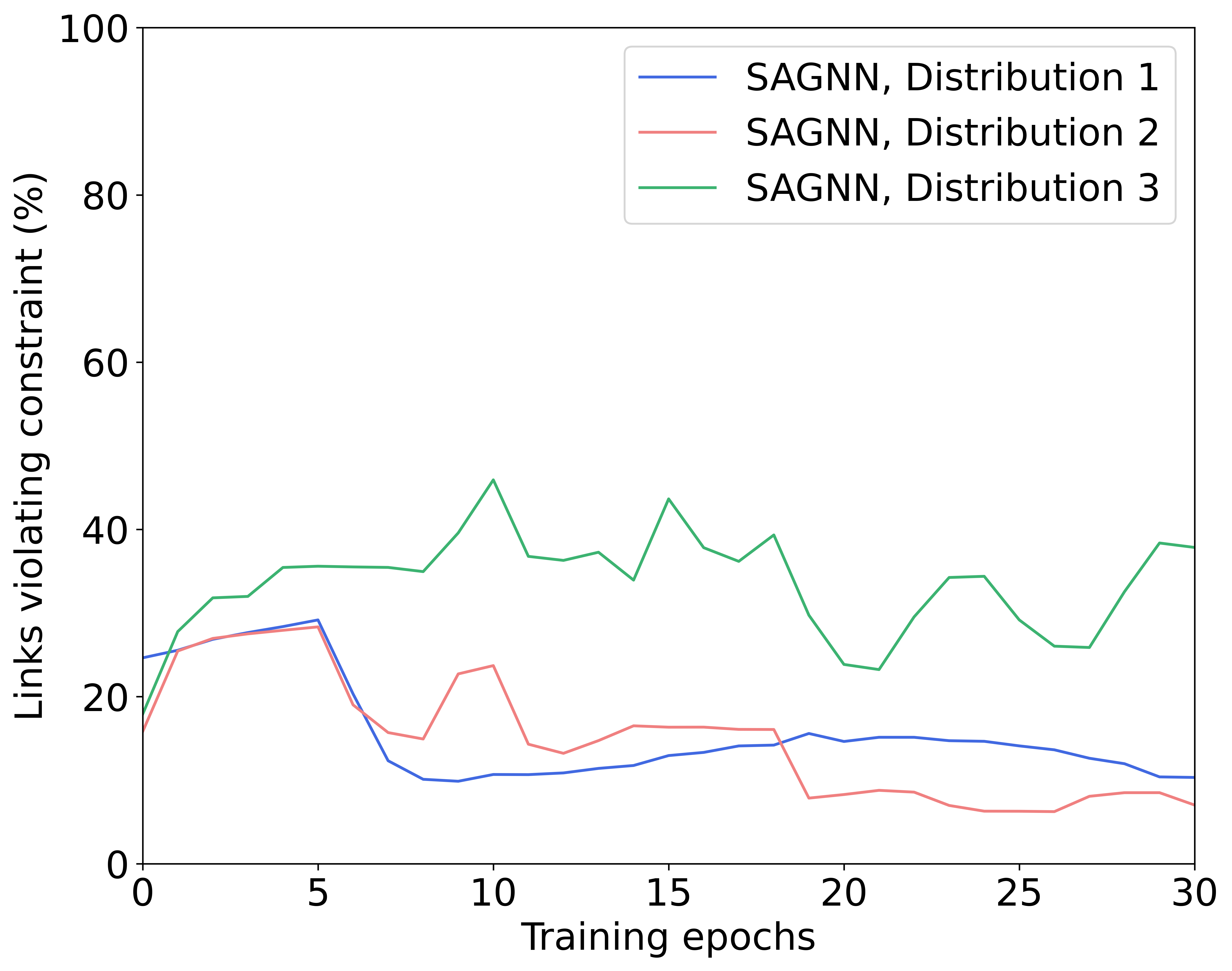}%
        \label{subfig:noise02}
    }
    \hfill
    \subfloat[High noise]{%
        \includegraphics[width=0.45\linewidth]{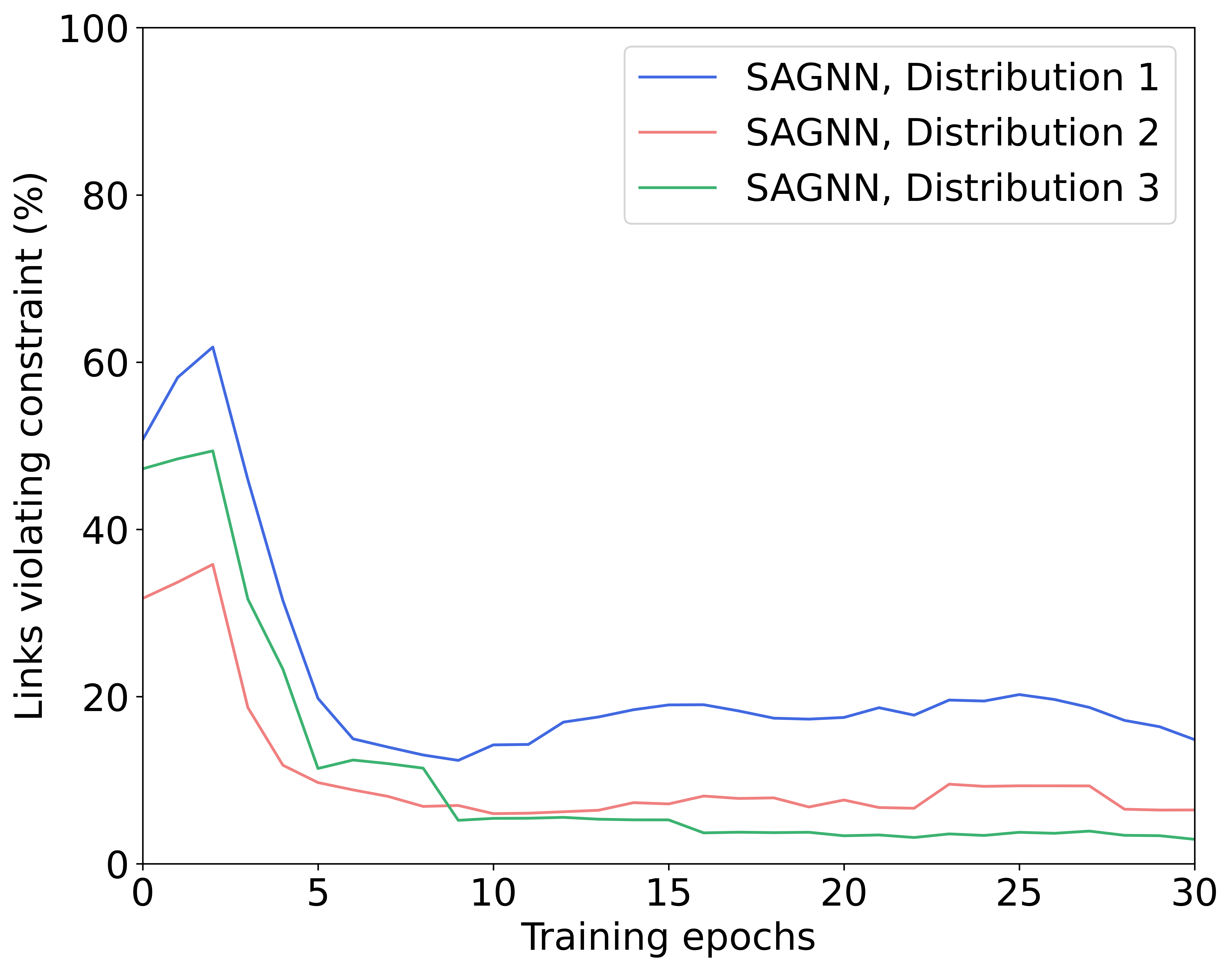}%
        \label{subfig:noise03}
    }
    \caption{Constraint violation evolution during training, running evaluation on the test dataset. We perform running averages on the time series to make them smoother.}
    \label{fig:cvdistributions}
\end{figure}

In Figure \ref{fig:cvdistributions}, we show the evolution of constraint violations over the first 30 training epochs for each dataset, evaluated under different sampling strategies for $\bm{\lambda}$. Distribution 1 samples each entry of $\bm{\lambda}$ uniformly between 0 and a fixed scale (set to 2). To expose the model to edge cases, a fixed percentage of entries are manually set to 0 or 1 in some instances. Distribution 2 combines uniform sampling with empirical sampling: half of the dual variable vectors are sampled uniformly, while the other half are drawn from a stored set of $\bm{\lambda}$ vectors collected during extra evaluation runs on the training graphs. Distribution 3 emphasizes reuse of previous training signals: uniform sampling occurs in only $10\%$ of iterations, while the remaining $90\%$ sample from the stored empirical set of dual variables. These strategies allow the model to train under a mix of exploratory and realistic scheduling demands. As shown in Figure \ref{fig:cvdistributions}, increased graph irregularity leads to better performance when a greater proportion of previously seen $\bm{\lambda}$ vectors are reused. This aligns with the intuition that, as the degree distribution deviates further from uniformity, purely random sampling becomes less effective at capturing the structure needed for constraint satisfaction.

 %%%%%%%%%%%%%%%%%%%%%%%%%%%%%%%%%%%%%%%%%%%%%%%%%%%%%%%%%%%%%%%%%%%%%%%%%%%%%%%%
%%%   APPENDIX   %%%%%%%%%%%%%%%%%%%%%%%%%%%%%%%%%%%%%%%%%%%%%%%%%%%%%%%%%%%%%%%
%%%%%%%%%%%%%%%%%%%%%%%%%%%%%%%%%%%%%%%%%%%%%%%%%%%%%%%%%%%%%%%%%%%%%%%%%%%%%%%%

 \label{sec:append}

% \clearpage
% \setcounter{page}{1}
% \begin{center}
% \textbf{\large Supplemental Materials}
% \end{center}

% \section{Supplementary Materials}
% \input{suplement}

\end{document}